\documentclass[aps,prd,final,reprint,showkeys,showpacs,superscriptaddress,floatfix,nofootinbib,twocolumn,amsmath,amssymb]{revtex4-1}

\usepackage[utf8]{inputenc}

\usepackage{hyperref} 
\usepackage{url}
\usepackage{graphicx}

\usepackage{graphics}

\graphicspath{{./}}

\usepackage[normalem]{ulem}
\usepackage{color}

\usepackage{soul}

\def\units#1{~\hbox{$\,{\rm #1}$}}

\begin{document}

\title{Search for Features in the Cosmic-Ray Electron and Positron spectrum measured by the Fermi Large Area Telescope}

\author{M.~N.~Mazziotta}
\email{mazziotta@ba.infn.it}
\affiliation{Istituto Nazionale di Fisica Nucleare, Sezione di Bari, via Orabona 4, I-70126 Bari, Italy}
\author{F.~Costanza}
\affiliation{Istituto Nazionale di Fisica Nucleare, Sezione di Bari, via Orabona 4, I-70126 Bari, Italy}
\affiliation{CNRS - Laboratoire d’Annecy de Physique des Particules, 9 Chemin de Bellevue, F-74940 Annecy, France}
\author{A.~Cuoco}
\affiliation{RWTH Aachen University, Institute for Theoretical Particle Physics and Cosmology (TTK), D-52056 Aachen, Germany}
\affiliation{Istituto Nazionale di Fisica Nucleare, Sezione di Torino, via Pietro Giuria 1,  I-10125 Torino, Italy}
\author{F.~Gargano}
\affiliation{Istituto Nazionale di Fisica Nucleare, Sezione di Bari, via Orabona 4, I-70126 Bari, Italy}
\author{F.~Loparco}
\affiliation{Istituto Nazionale di Fisica Nucleare, Sezione di Bari, via Orabona 4, I-70126 Bari, Italy}
\affiliation{Dipartimento di Fisica ``M. Merlin" dell'Universit\`a e del Politecnico di Bari, via Amendola 173, I-70126 Bari, Italy}
\author{S.~Zimmer}
\affiliation{University of Geneva, D\'epartement de physique nucl\'eaire et corpusculaire (DPNC), 24 quai Ernest-Ansermet, CH-1211 Gen\`eve 4, Switzerland}

\date{\today}

\begin{abstract}
The Large Area Telescope onboard the Fermi Gamma-ray Space Telescope has collected the
largest ever sample of high-energy cosmic-ray electron and positron events. 
Possible features in their energy spectrum could be a signature of the presence
of nearby astrophysical sources, or of more exotic sources, such as annihilation or decay 
of dark matter (DM) particles in the Galaxy.
In this paper for the first time we search for a delta-like line feature in
the cosmic-ray electron and positron spectrum. We also search for a possible feature
originating from DM particles annihilating into electron-positron pairs.
Both searches yield negative results, but we are able to set constraints on the 
line intensity and on the velocity-averaged DM annihilation cross section.
Our limits extend up to DM masses of 1.7$\units{TeV/c^2}$, and exclude the thermal value
of the annihilation cross-section for DM lighter than 150$\units{GeV/c^2}$.
\end{abstract}



\maketitle

\section{Introduction}
\label{intro}
During their propagation in our Galaxy, high-energy cosmic-ray electrons and positrons 
(CREs) lose their energy mainly through synchrotron radiation and inverse Compton interactions 
with the low-energy photons of the interstellar radiation field. Therefore, CREs reaching 
the Earth with energies above $100 \units{GeV}$ should be produced by a few nearby 
sources~\cite{Ackermann:2010ip,Abdollahi:2017kyf}.
Searching for anisotropies in the CRE spectrum provides a powerful probe for local sources,
but current limits strongly disfavor the presence of local young and middle-aged
astrophysical sources since such sources would produce large anisotropies~\cite{Abdollahi:2017kyf}. 

An alternative production mechanism for high-energy CREs could arise due to
the annihilation or decay of dark matter (DM), which would yield 
anisotropies in the CRE flux, albeit below the sensitivity of current 
analyses~\cite{Borriello:2010qh}. 
In this case, the CRE energy spectrum is expected to exhibit a
cut-off at the energy corresponding to the DM 
mass~\cite{Cirelli:2010xx}. This feature will still be visible 
in the spectrum after propagation. Therefore, the signature of a DM contribution 
to the CRE spectrum would be an ``edge''-like feature at energies close to the DM mass.

Further features in the spectrum are expected from the fact that only a few astrophysical sources
will contribute at the highest energies. 
In fact, CRE spectra of pulsars or supernova remnants are expected to be power laws
with cut-offs, which vary from source to source. Thus the superposition 
of different sources will produce a final spectrum with bumps and dips, 
which will be more pronounced the fewer sources contribute~\cite{Grasso:2009ma}.
The cut-offs are expected to be softer for these sources, with  respect to the DM case,
although, in practice, in the presence of a weak signal, it would be difficult to 
distinguish the two.

The CRE energy spectrum has been measured by several experiments, 
like AMS-02~\cite{Aguilar:2014fea}, CALET~\cite{Adriani:2017efm}, 
DAMPE~\cite{Ambrosi:2017wek} and the Fermi Large Area Telescope (LAT)~\cite{Abdo:2009zk,FermiLAT:2011ab,Abdollahi:2017nat}.
Recently, the results of new measurements have been published, 
providing a confirmation of a break in the $\units{TeV}$ region as previously seen by ground-based Cherenkov gamma-ray telescopes~\cite{Aharonian:2008aa,Aharonian:2009ah}
and an indication of a potential feature at $1.4 \units{TeV}$ (DAMPE).

The CRE data from the Large Area Telescope (LAT) onboard
the Fermi satellite~\cite{Atwood:2009ez,Abdo:2009gy,Ackermann:2012kna} 
have already been used to measure the energy 
spectrum~\cite{Abdo:2009zk,FermiLAT:2011ab,Abdollahi:2017nat}, to search for 
anisotropies~\cite{Ackermann:2010ip,Abdollahi:2017kyf} and for
a possible excess from the Sun~\cite{Ajello:2011dq}. 
In this paper we have analyzed the same data sample used 
in the measurement of the CRE energy spectrum reported in Ref.~\cite{Abdollahi:2017nat}
for the high-energy analysis (standard path-length selection).\footnote{The 
data set has been collected between August 4, 2008, and June 24, 2015. 
The full details of the event selection are reported in Ref.~\cite{Abdollahi:2017nat}.}

In this analysis for the first time we use the Fermi-LAT CRE data to
search for possible features in the spectrum originating from the 
direct annihilation of DM particles into $e^{+}e^{-}$ pairs. 
In particular, as will be illustrated in Sec.~\ref{anamet}, 
we will search for either delta-like lines or for spectral edges.
In the past, several attempts~\cite{DiMauro:2015jxa,Bergstrom:2009fa,Feng:2012gs} 
were made to constrain scenarios with DM particles annihilating or decaying to 
leptonic final states using the measurements of CRE spectra performed by various experiments.
In particular, in Refs.~\cite{Bergstrom:2013jra,Cavasonza:2016qem} 
the CR positron ratio and the separate positron and electron fluxes measured by the AMS-02 experiment were used
to constrain direct DM annihilations into $e^{+}e^{-}$.
The LAT CRE data extend to higher energies than those from AMS-02, 
thus allowing us to set constraints for higher DM masses.

\section{Spectrum of Cosmic-ray electrons and positrons from DM annihilations}
\label{DMprop}
To evaluate the spectrum of CREs produced from DM annihilations in the Galaxy, we used a customized 
version of the propagation code {\tt DRAGON}~\cite{Evoli:2008dv,Gaggero:2013rya,dragonweb},
in which the cross sections for the production of secondary particles are taken from Ref.~\cite{Mazziotta:2015uba}.
We set the propagation model of CRs in the Galaxy assuming the
source term distribution from Ref.~\cite{Ferriere:2001rg}, while
the gas density distribution and the interstellar radiation field (ISRF) 
are taken from the public {\tt GALPROP} version~\cite{Strong:1998pw,Moskalenko:2001ya,galpropweb}.
The Galactic magnetic field model (GMF) is taken from Ref.~\cite{Pshirkov:2011um}.

We adopted a 3D version of the {\tt DRAGON} code including a spiral arms model~\cite{SteimanCameron:2010tm}
that superimposes the spatial pattern of the distribution of different astrophysical quantities (e.g. source term,
gas, ISRF and magnetic field)~\cite{Evoli:2016xgn}. In our simulation we assume that the interstellar 
medium is composed of Hydrogen and Helium with relative abundances $1:0.1$.

\begin{figure}[!ht]
\begin{center}
\includegraphics[width=\columnwidth,height=0.22\textheight,clip]{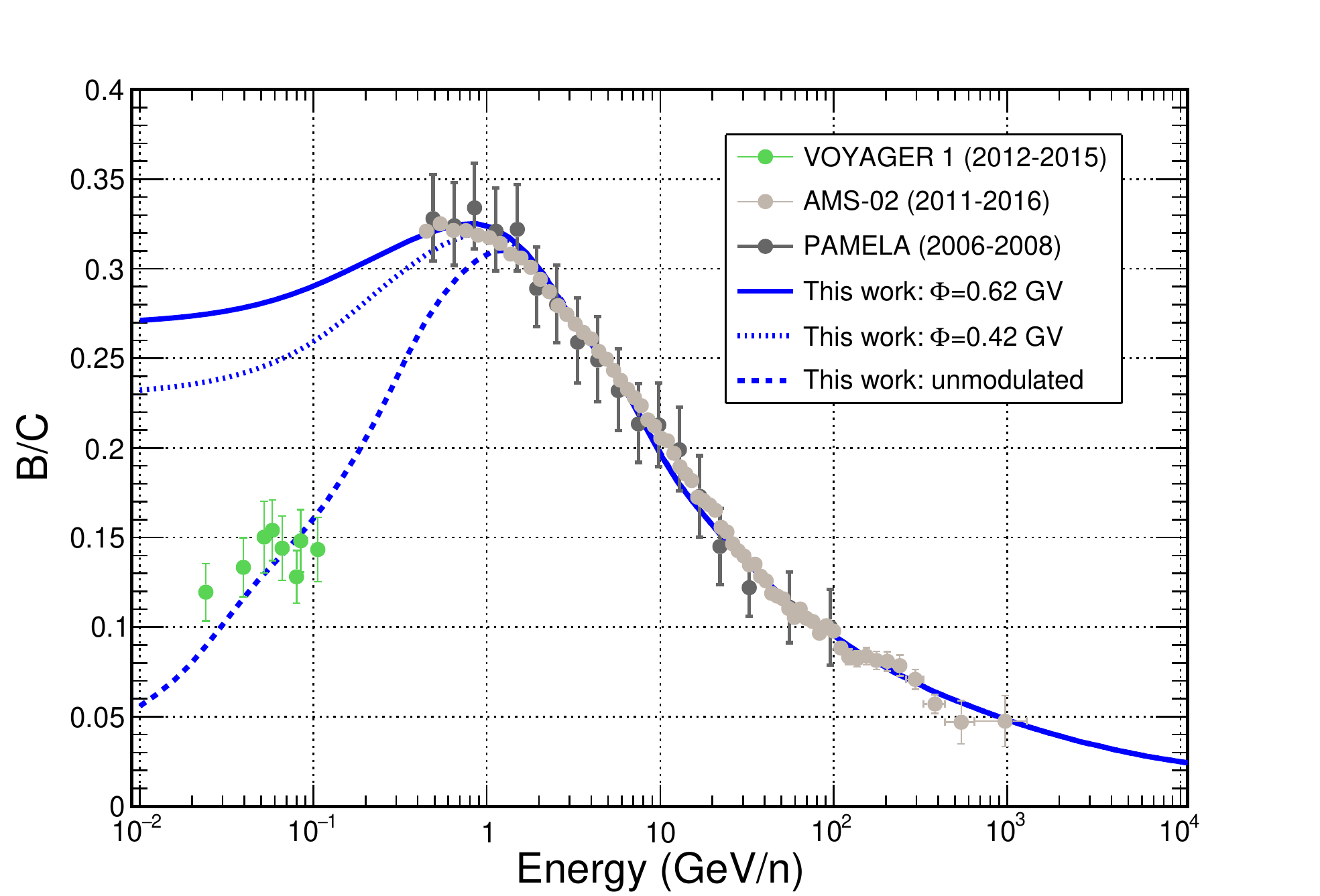}
\end{center}
\caption{Comparison of the predictions from the propagation model with boron to carbon ratio (B/C) data observed near the Earth.
Dashed line: unmodulated intensity; dotted 
(solid) line: modulated intensity by means of the force field approximation with $\Phi = 0.42~(0.62) \units{GV}$,
respectively. The plots show the data from VOYAGER 1~\cite{Stone150,0004-637X-831-1-18}, PAMELA~\cite{2014ApJ...791...93A}
and AMS-02~\cite{Aguilar:2016vqr}.}
\label{FigBC}
\end{figure}

We assume that the scalar diffusion coefficient depends on the particle rigidity $R$ and on the distance from 
the Galactic plane $z$ according to the parametrization 
$D \,\, = \,\, D_0 \, \beta^{\eta} \, \left( R / R_0 \right)^{\alpha} \, e^{|z|/z_t}$
~\cite{Gaggero:2013nfa}.
We set $\alpha=0.33$ according to the recent boron to carbon ratio (B/C) from the AMS-02 data~\cite{Aguilar:2016vqr}, $R_0 = 4 \units{GV}$ 
and $z_t = 4 \units{kpc}$, 
while $D_0$ and $\eta$ are tuned to the B/C AMS-02 data, also setting the nuclei injection spectra to reproduce
the \mbox{VOYAGER 1} data at low energy~\cite{Stone150,0004-637X-831-1-18}. We have found that the B/C data are reproduced setting 
$D_0=7.4 \times 10^{28} \units{cm^2~s^{-1}}$ and $\eta=-0.1$. 
A reacceleration model is also adopted to reproduce the B/C data at low energy, setting
the Alfv\'en velocity to $v_{\rm A} = 52 \units{km~s^{-1}}$.
The solar modulation is treated using the force-field approximation~\cite{Gleeson:1968zza} 
with $\Phi = 0.42\units{GV}$ and $\Phi = 0.62\units{GV}$ to reproduce the PAMELA and AMS-02 data respectively,
which were taken at different parts of the solar cycle.

Figure~\ref{FigBC} shows the comparison of the predictions from the actual propagation model 
with B/C data observed near the Earth by PAMELA~\cite{2014ApJ...791...93A} and by AMS-02~\cite{Aguilar:2016vqr}, and outside
the solar system by VOYAGER 1~\cite{Stone150,0004-637X-831-1-18} (unmodulated model).

\begin{figure}[!ht]
\begin{center}
\includegraphics[width=\columnwidth,height=0.22\textheight,clip]{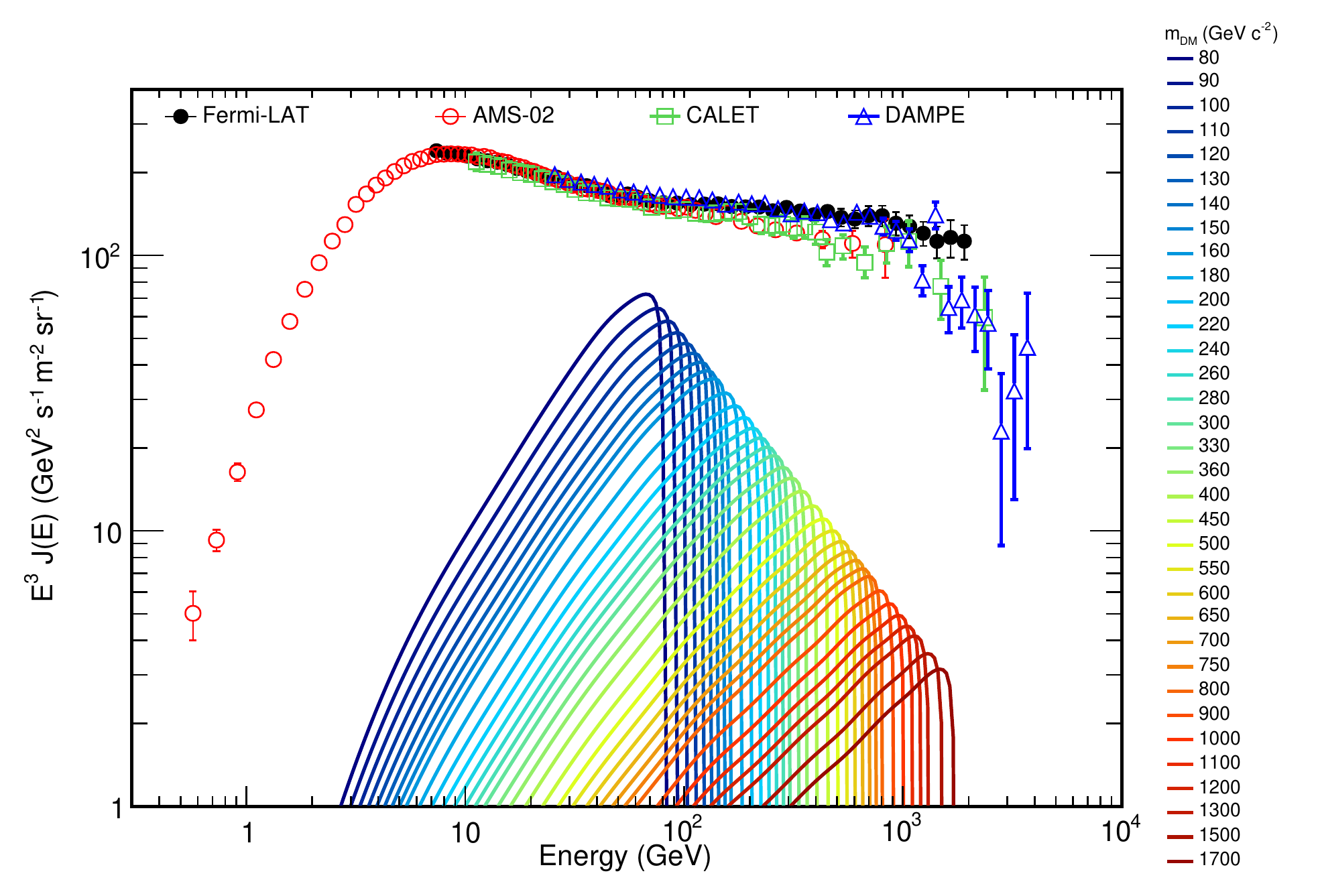}
\end{center}
\caption{
CRE spectra at Earth. The expected contributions from possible DM annihilations in the Galaxy (see text)
are compared with the data from the Fermi LAT~\cite{Abdollahi:2017nat}, AMS-02~\cite{Aguilar:2014fea}, 
CALET~\cite{Adriani:2017efm} and DAMPE~\cite{Ambrosi:2017wek}.}
\label{FigDMCRE}
\end{figure}

We have used our model to propagate the CREs produced by DM annihilations in our Galaxy using the
{\tt DRAGON} code. We assume a 
Navarro–Frenk–White (NFW) DM density profile~\cite{1996ApJ...462..563N} with a local DM density 
$\rho_\odot=0.41~\units{GeV~cm^{-3}}$~\cite{Salucci:2010qr}, and an annihilation cross section 
$\langle \rm{\sigma v} \rangle = 3 \times 10^{-26} \units{cm^3~s^{-1}}$.
The inclusive yields of $e^{\pm}$ from DM annihilations are taken from~\cite{Cirelli:2010xx}, 
including electroweak corrections~\cite{1475-7516-2011-03-019}.
Fig.~\ref{FigDMCRE} shows, for each DM mass, the expected CRE spectra at Earth 
(scaled by a factor of 10, i.e. they correspond to $\langle \rm{\sigma v} \rangle = 3 \times 10^{-25} \units{cm^3~s^{-1}}$)
compared with the Fermi LAT~\cite{Abdollahi:2017nat}, AMS02~\cite{Aguilar:2014fea},
CALET~\cite{Adriani:2017efm} and DAMPE~\cite{Ambrosi:2017wek} data.
The DM spectra have been modulated using the force-field approximation with a
modulation potential $\Phi = 0.55\units{GV}$. This value has been derived based on an analysis of gamma 
rays coming from the Moon using the same time range as considered in this paper
~\cite{Cerutti:2016gts}. The DM spectra are used as templates in the fit procedure described in Sec.~\ref{anamet}.
For different $\rho_\odot$, constraints will rescale as $\rho_\odot^2$.

We have also used the 2D version of {\tt DRAGON}~\cite{Evoli:2016xgn}
in which the diffuse equation is solved in cylindrical coordinates 
with azimuthal symmetry and without spiral arms. In this way we can check for the effects of uncertainties related to 
propagation on the DM spectra. We used the same values 
of $R_0$, $z_t$ as in the 3D model, while the remaining parameters 
have been adjusted to the B/C data. In particular we found $\alpha=0.42$, resulting in DM spectra lower by about $20\%$ or 
less with respect to those evaluated with the 3D model of the Galaxy.
We have tested also 2D models with different $z_t$ from $2 \units{kpc}$ to $7 \units{kpc}$.
In this case the effect is smaller, at the level of $10\%$.

Further uncertainties come from the ISRF and the GMF.
We have studied them changing separately the normalization of the ISRF and magnetic field by $\pm 50\%$, 
which resulted in a $[-50\%, +100\%]$ variation in the normalization of DM spectra.
A similar study has been performed in Ref.~\cite{Bergstrom:2013jra}
yielding comparable results.

We have also tested for a different DM profile, namely an isothermal profile,
still normalized to the same local DM density. Differences in this case
are even smaller (few percent).

\section{Analysis Method}
\label{anamet}
Following the approach of Ref.~\cite{Ackermann:2015lka}, we have implemented a  
fitting procedure in sliding energy windows to search for possible local peaks 
(either bumps or lines) on top of a smooth CRE spectrum. 

In each energy window we model the CRE intensity as $I(E) = I_{0}(E) + I_{f}(E)$,
where $I_{0}(E)$ is the ``smooth'' part of the spectrum and $I_{f}(E)$ describes 
the possible feature. Since the energy windows are narrow, we assume that the smooth 
part of the spectrum can be described by a power-law (PL) model 
$I_{0}(E) = k  (E/E_{0})^{-\gamma}$,
where $\gamma$ is the PL spectral index and the prefactor $k$ corresponds the CRE intensity at 
the scale energy $E_{0}$, fixed to $1 \units{GeV}$.

In our analysis, we assume two models for $I_{f}(E)$:
(i) a delta-like (line) model $I_{f}(E) = s  \delta (E - E_{line})$, where $s$ represents the line intensity;
(ii) a spectrum produced by DM annihilating into CREs 
$I_{f}(E) = s I_{DM}(E | m_{DM}, \langle \rm{\sigma v} \rangle, ...)$,
where $I_{DM}(E)$ is the intensity of CREs from DM observed near the Earth, which is calculated in Sec.~\ref{DMprop}, 
and the parameter $s$ represents the scale of the annihilation cross-section implemented in the model.
In this case $s$ corresponds to $\langle \rm{\sigma v} \rangle$ in units of $3 \times 10^{-26} \units{cm^3~s^{-1}}$.   
The line model is used as a generic model for a feature.
It can represent DM spectra from alternative DM models, or also features induced
by local nearby astrophysical sources.

Starting from the model, we can calculate the expected counts in each CRE observed energy bin $E_j$ as:

\begin{equation}
\mu_j = \mu(E_j) = t  \int dE ~ \mathcal{R}(E_j | E) ~ I(E)
\end{equation}
where $E$ is true (Monte Carlo) energy,   
$\mathcal{R}(E_j | E)$ is the instrument response matrix 
(acceptance) which incorporates the energy resolution 
of the LAT, and $t$ is the integrated livetime.

For our fitting procedure we define a $\chi^2$ function as follows:

\begin{equation}
  \chi^2 = \sum_{j=1}^{N} \frac{ \left( n_j - \mu_{j} \right) ^2}{ n_j + f^2_{syst} n^2_j}
   \label{eq:likelihood}
\end{equation}
where $N$ is the number of energy bins used for the fit. 
The denominator of each term in the summation includes the sum in quadrature of the statistical Poisson
fluctuations ($\sqrt{n_j}$) and systematic uncertainties ($f_{syst} n_j$), which are discussed
more in detail below.

To estimate the parameters $\{ k, \gamma, s\}$ which minimize the
$\chi^{2}$ we use the {\tt MINUIT} code within the 
ROOT toolkit~\cite{Brun:1997pa,rootweb};
the values of the parameters at a $95\%$ confidence limit (CL) 
are evaluated using {\tt MINOS} and setting the error confidence level to $2.71$.

\begin{figure*}[!ht]
\includegraphics[width=\columnwidth,height=0.28\textheight]{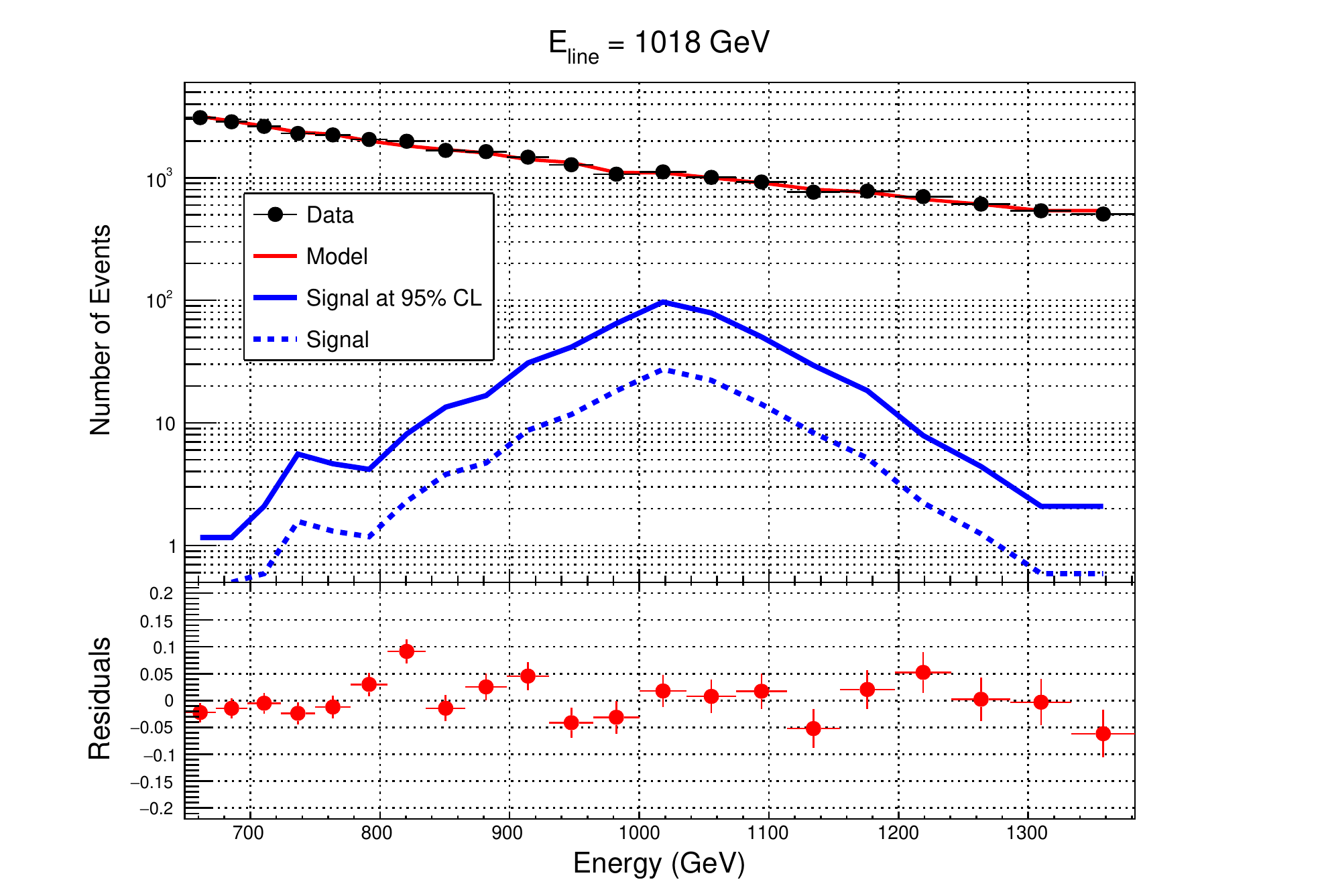} 
\includegraphics[width=\columnwidth,height=0.28\textheight]{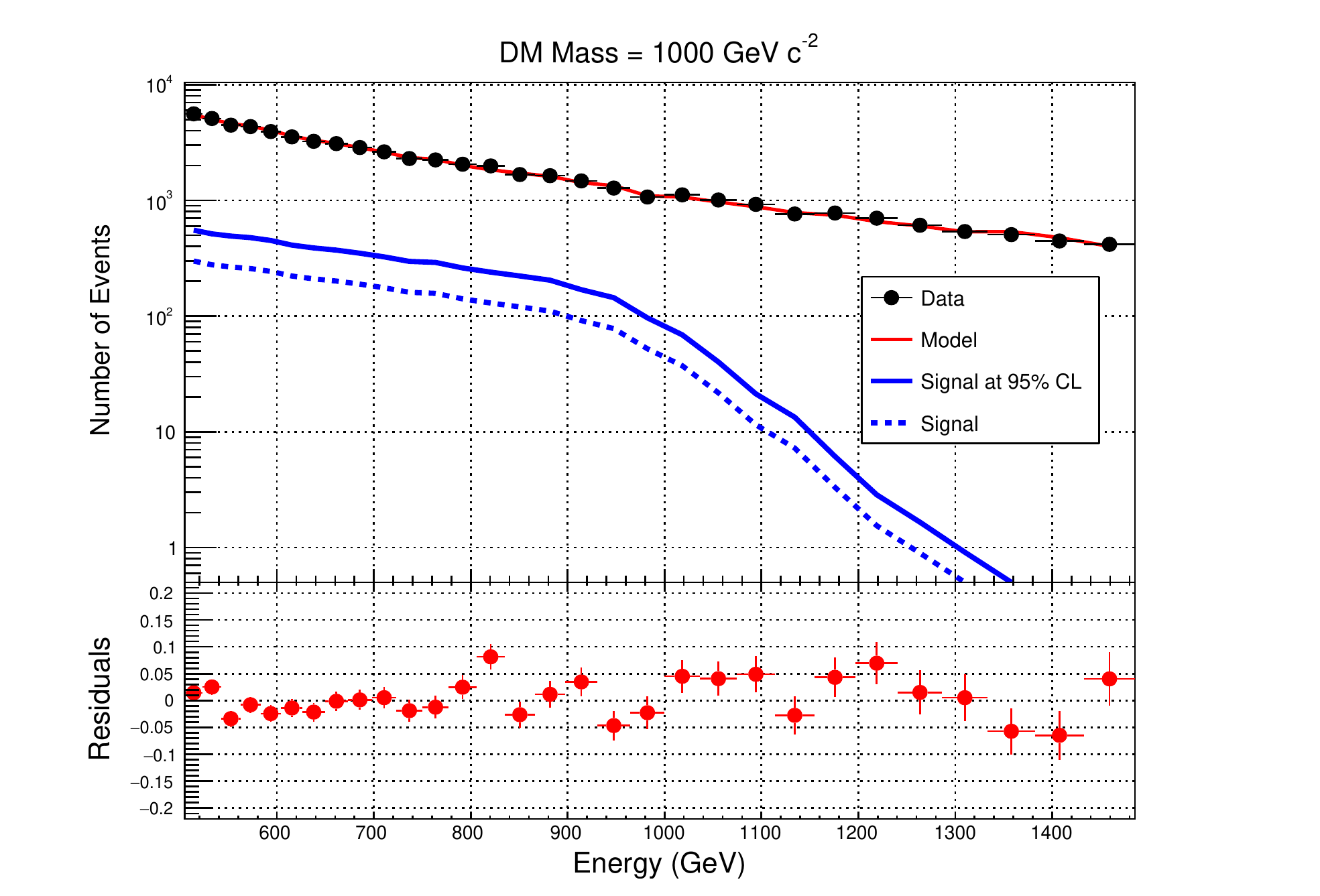} 
\caption{Example of fit results near $1 \units{TeV}$. The left plot is obtained 
fitting the CRE spectrum with a delta-like line feature on top of a
PL spectrum; the right plot is obtained assuming a feature due to DM 
annihilating into CREs on top of a PL spectrum. 
The top panels of each plot show a comparison of the measured counts (black points) 
with those predicted from the fit (red solid line). 
The contributions from the possible features are also shown:
the blue dashed lines indicate the counts originated from the feature when the best fit value
for the parameter $s$ is assumed; the blue solid lines indicate the counts originated from the 
feature when the upper limit at $95\%$ confidence level for $s$ is assumed. 
The bottom panels show the fit residuals as a function of energy. 
The error bars include only the statistical uncertainties.}
\label{Fig:exmfit}
\end{figure*}

We have scanned an energy range extending from $42 \units{GeV}$ to $2 \units{TeV}$.\footnote{The 
limits of $42 \units{GeV}$ and $2 \units{TeV}$ are the same as in the high-energy event 
selection in Ref.~\cite{Abdollahi:2017nat}.}
This interval has been divided in $64$ bins per decade, equally spaced on a logarithmic scale.
When searching for line features, we selected fit windows centered on the
line energy $E_{line}$ with a half-width of $0.35 E_{line}$.
Once folded with the energy response, a delta-like 
line will show up as a broad peak in the count spectrum with the same width as the energy resolution of the LAT,
which is always less than the window size\footnote{The 
LAT energy resolution for the CRE selection at $95\%$ containment 
ranges from about $15\%$ at $42 \units{GeV}$ to about $20\%$ at $1 \units{TeV}$
and increases up to $35\%$ at $2 \units{TeV}$~\cite{Abdollahi:2017nat}.}.
On the other hand, when searching for a DM signal, we selected fit windows centered on the 
candidate DM mass $m_{DM}$ with a half-width of $0.5 m_{DM}$.
Since a feature originating from DM (Sec.~\ref{DMprop})
will be spread across a larger energy interval than a line,
we chose a larger fit window than in the line search.
We also tested different energy binnings and 
different window sizes yielding comparable results.
The details of these studies are given in Appendix~\ref{sec:binning}.

The high-energy CRE sample used in the present analysis is affected by 
systematic uncertainties. In Ref.~\cite{Abdollahi:2017nat} it was shown that the
fractional systematic uncertainty $f_{syst}$, due to the acceptance calculation, 
to the proton contamination and to the data/Monte Carlo corrections (added in quadrature)
ranges from about $1.3\%$ at $42 \units{GeV}$ to about $15\%$ at $2 \units{TeV}$.
In Ref.~\cite{Abdollahi:2017nat} the calculation of $f_{syst}$ was
performed dividing the energy interval in $16$ bins per decade and 
the statistical uncertainties were always found to be about one of order of magnitude 
less than systematic ones.

To account for systematic uncertainties, that 
might mimic a false local feature signal or might mask a true local feature,
we have implemented a data-driven procedure.\footnote{Since 
no control measurements are available, we cannot evaluate systematic uncertainties  
following an approach like the one used in Ref.~\cite{Ackermann:2015lka}.}
As a starting point we have fitted the data, in a given window, with a PL 
model considering statistical uncertainties only. Then we have evaluated the
fractional residuals $f_j=(n_j-\mu_j)/\mu_j$, where $n_{j}$ is the number of CRE events 
in the $j$-th observed energy bin ($E_j$) and $\mu_{j}$ is the number of CRE events
predicted by the PL model. We have then built the distribution of fractional
residuals and we have calculated its root mean square (RMS).
Finally, we have derived the systematic uncertainties $f_{syst}$ from 
the difference between the observed RMS and its expected value 
when only statistical uncertainties are considered.\footnote{The 
RMS on the distribution of fractional residuals can be expressed as
$f_{RMS}^{2}=f_{stat}^{2}+f_{syst}^{2}$.}
We note that this is expected to slightly reduce the sensitivity to a possible
spectral feature, since the feature would contribute to the evaluation of the systematic uncertainties.
In the case of a non-detection, this would result in conservative limits.

\begin{figure*}[!ht]
\begin{tabular}{cc}
\includegraphics[width=\columnwidth,height=0.22\textheight,clip]{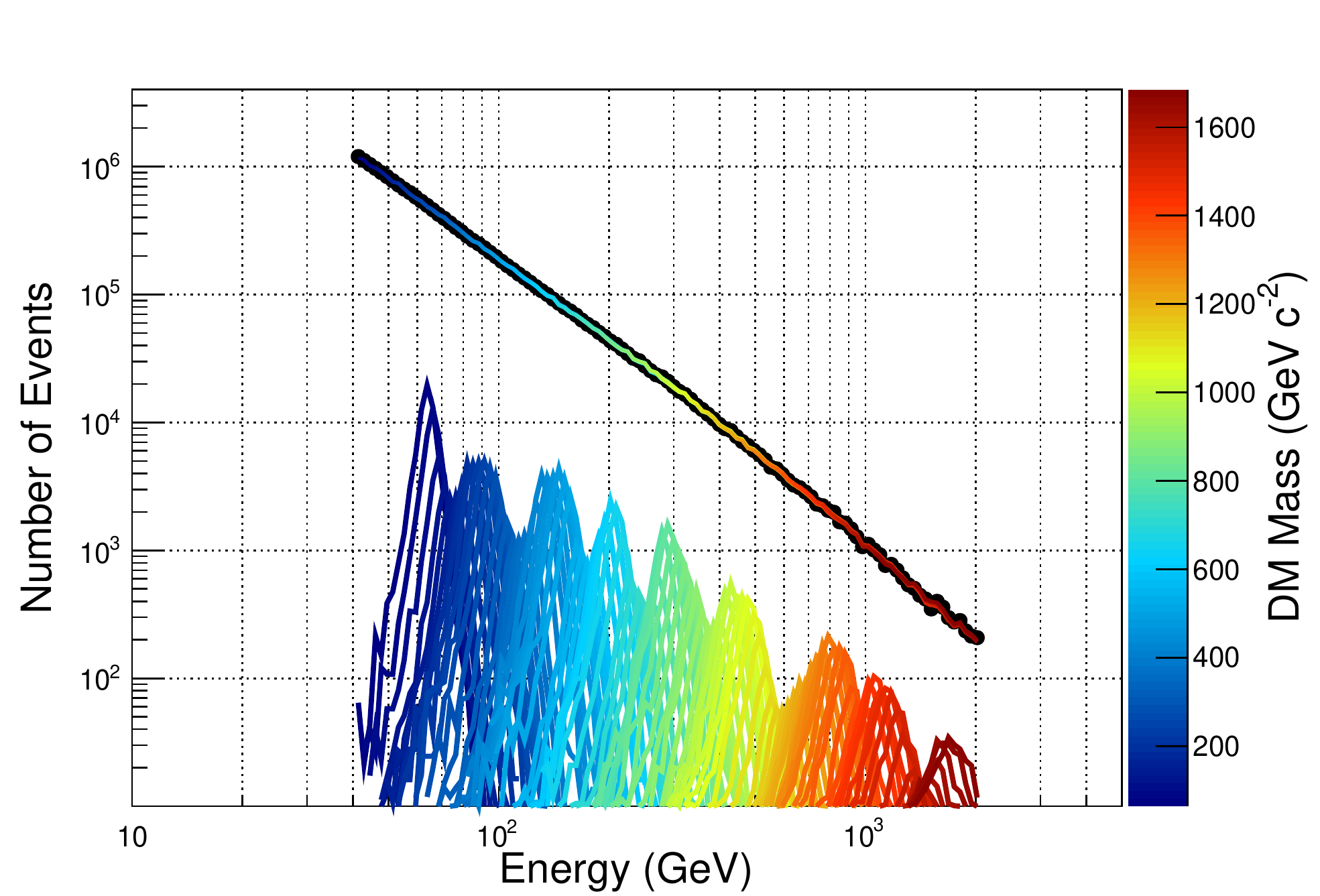} &
\includegraphics[width=\columnwidth,height=0.22\textheight,clip]{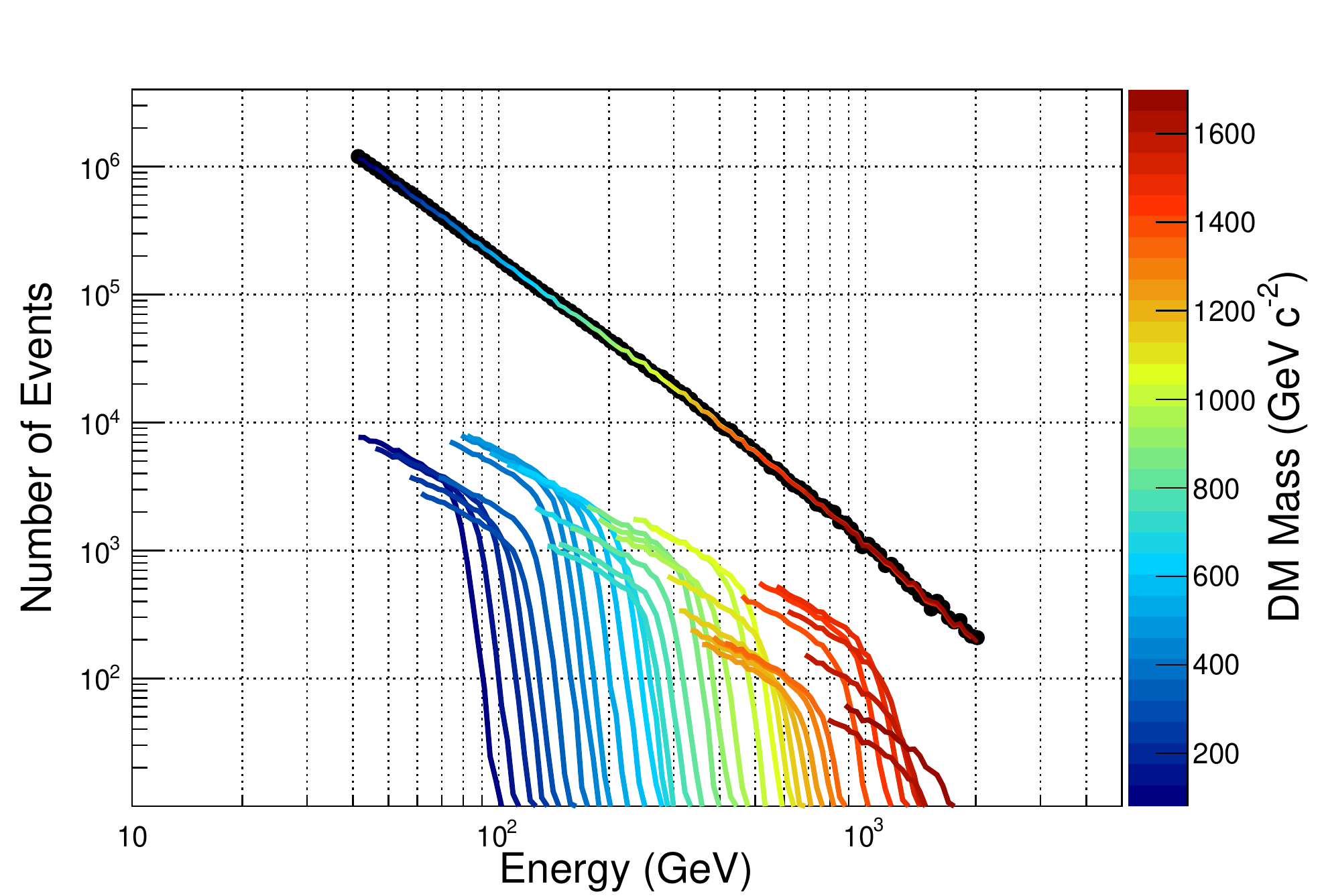} \\
\includegraphics[width=\columnwidth,height=0.22\textheight,clip]{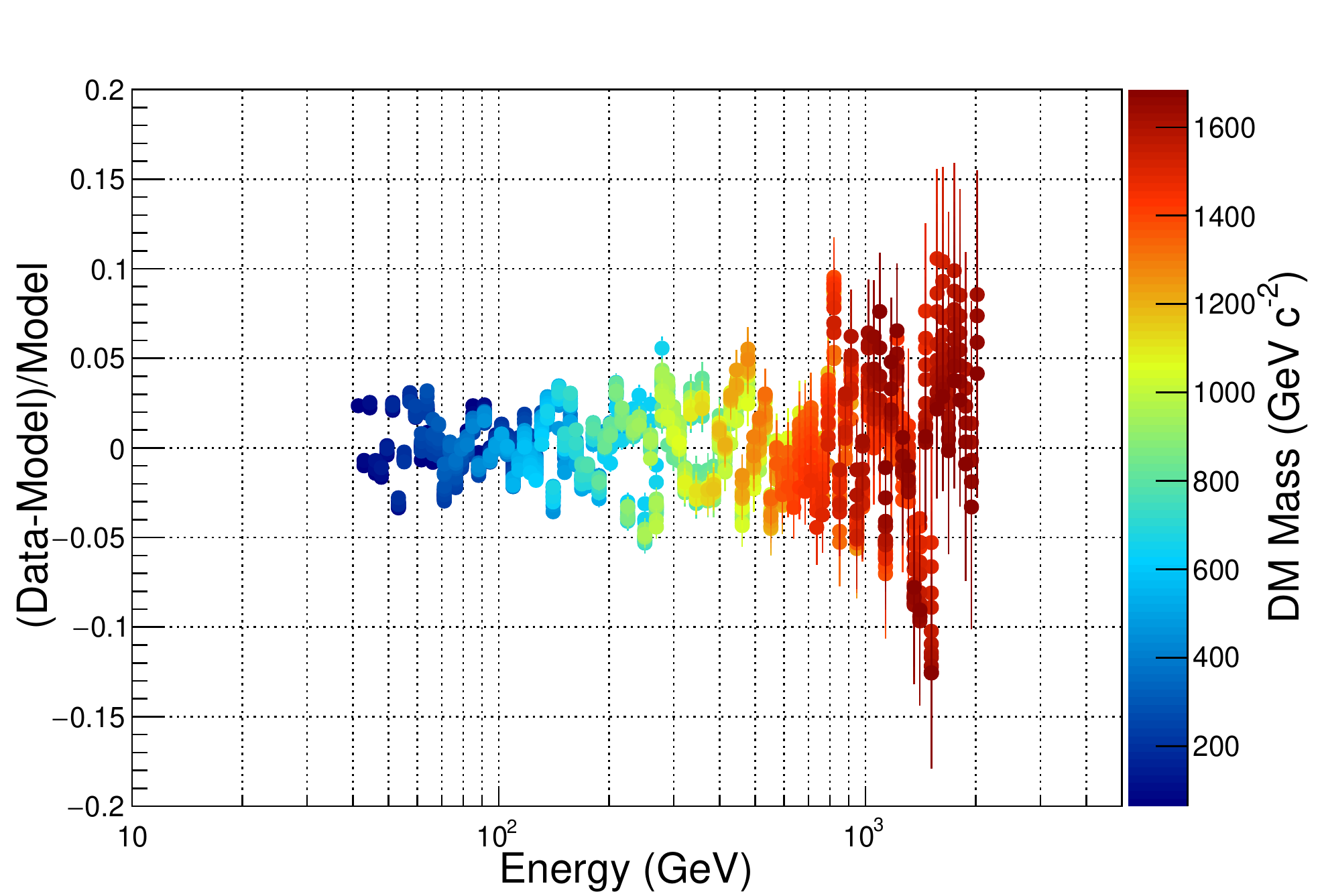} &
\includegraphics[width=\columnwidth,height=0.22\textheight,clip]{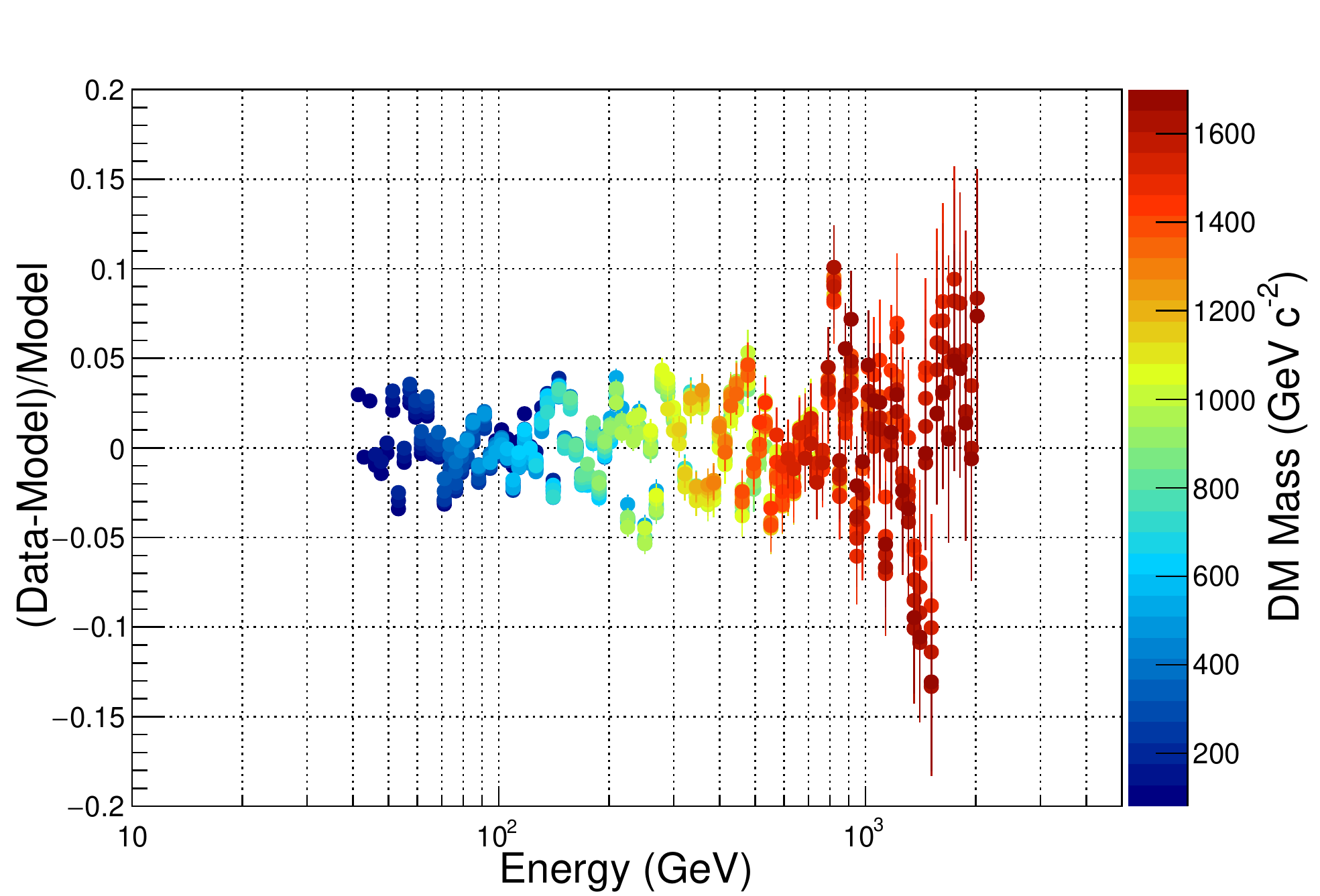} 
\end{tabular}
\caption{ 
Comparison of the fitted CRE spectra with the data.
The left plots have been obtained fitting the CRE 
spectrum with a delta-like line feature on top of a PL spectrum; the right plots 
have been obtained assuming a feature due to DM annihilating into CREs on top of a PL spectrum. 
Different colors correspond to different values of $E_{line}$ or $m_{DM}$ and, 
consequently, to different energy windows. The top plots show a comparison of the measured count 
spectra (black points) with the fitted ones (colored bands). The contributions from the features 
at $95\%$ CL limit are also shown in the plots as continuous solid lines. 
The bottom plots show the count residuals in the various energy windows. 
The error bars include only the statistical uncertainties.}
\label{Fig:results}
\end{figure*}

For each energy window we evaluate the significance of a possible feature 
considering the $\chi^2$ difference between the alternative hypothesis 
(line or DM signal) and the null hypothesis (PL model) as Test Statistics. 
In addition we evaluate the expectation bands for our results, i.e. the sensitivity to the null hypothesis, 
using a pseudo-experiment technique. As a starting point,
we fit the observed CRE count distribution with a simple PL model 
in the whole energy range.\footnote{As shown in Ref.~\cite{Abdollahi:2017nat}, the CRE energy
spectrum above $50 \units{GeV}$ is well described by a single power law.} 
This model is used as a template to evaluate the expected counts in each energy bin.
Starting from the template model, a set of 1000 pseudo-experiments is performed, in which
the counts in each energy bin are extracted from a Poisson distribution with mean
value taken from the template, after adding a gaussian fluctuation to account
for energy-dependent systematic uncertainties.    
The count distributions corresponding to the various pseudo-experiments are then
fitted including the feature, and the containment bands (quantiles) for all the parameters 
are calculated. 

\begin{figure*}[!ht]
\begin{tabular}{cc}
\includegraphics[width=\columnwidth,height=0.22\textheight,clip]{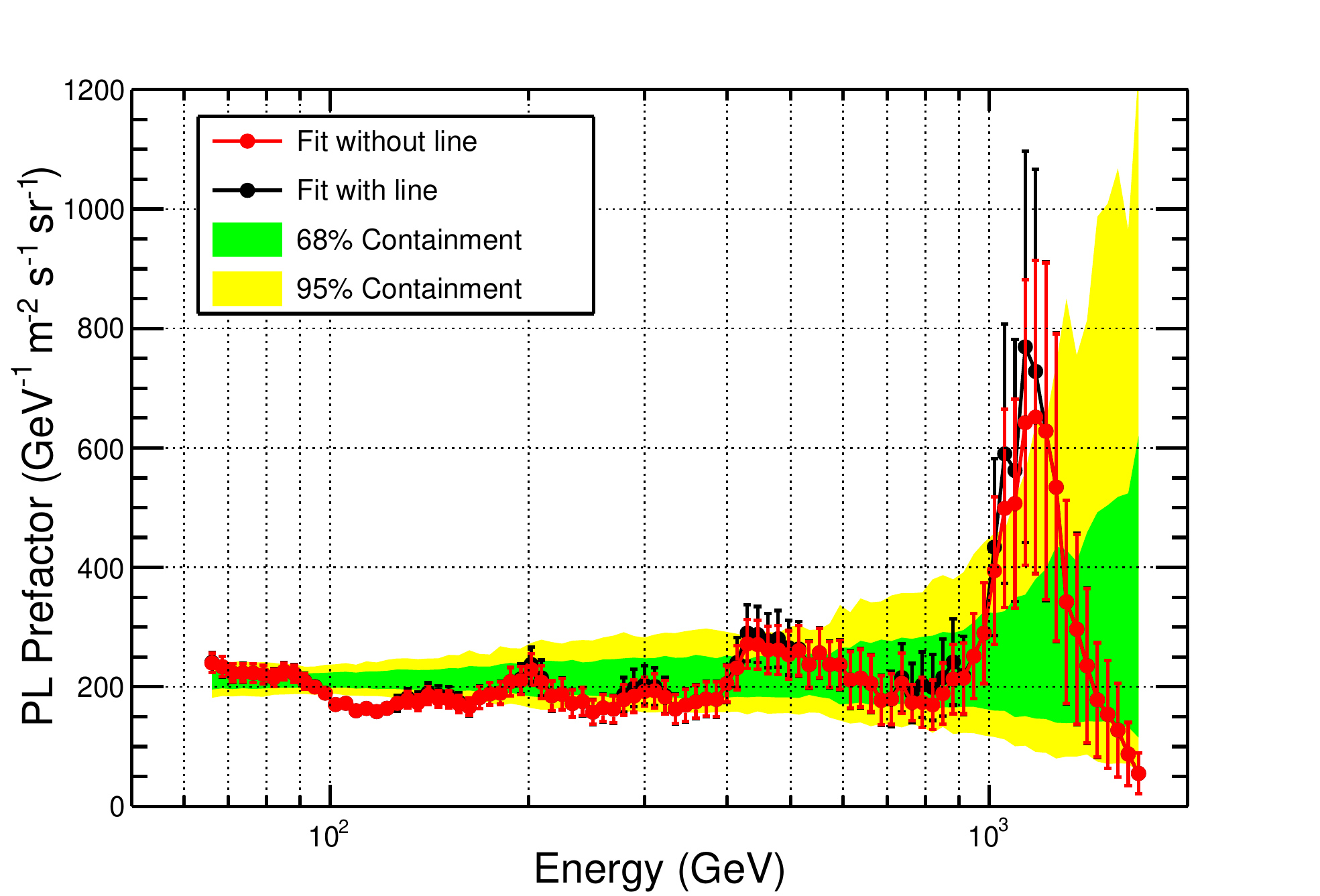} &
\includegraphics[width=\columnwidth,height=0.22\textheight,clip]{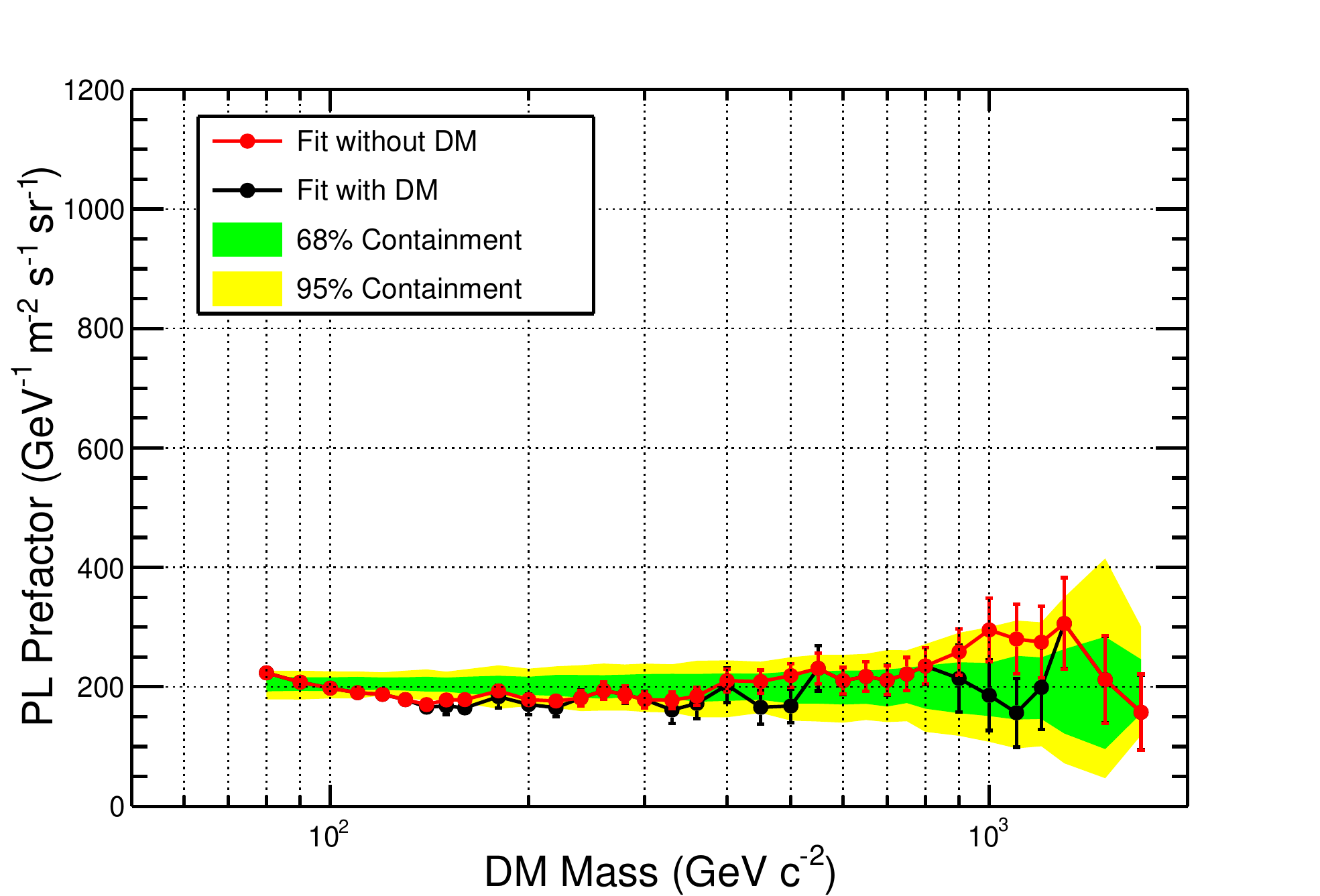} \\
\includegraphics[width=\columnwidth,height=0.22\textheight,clip]{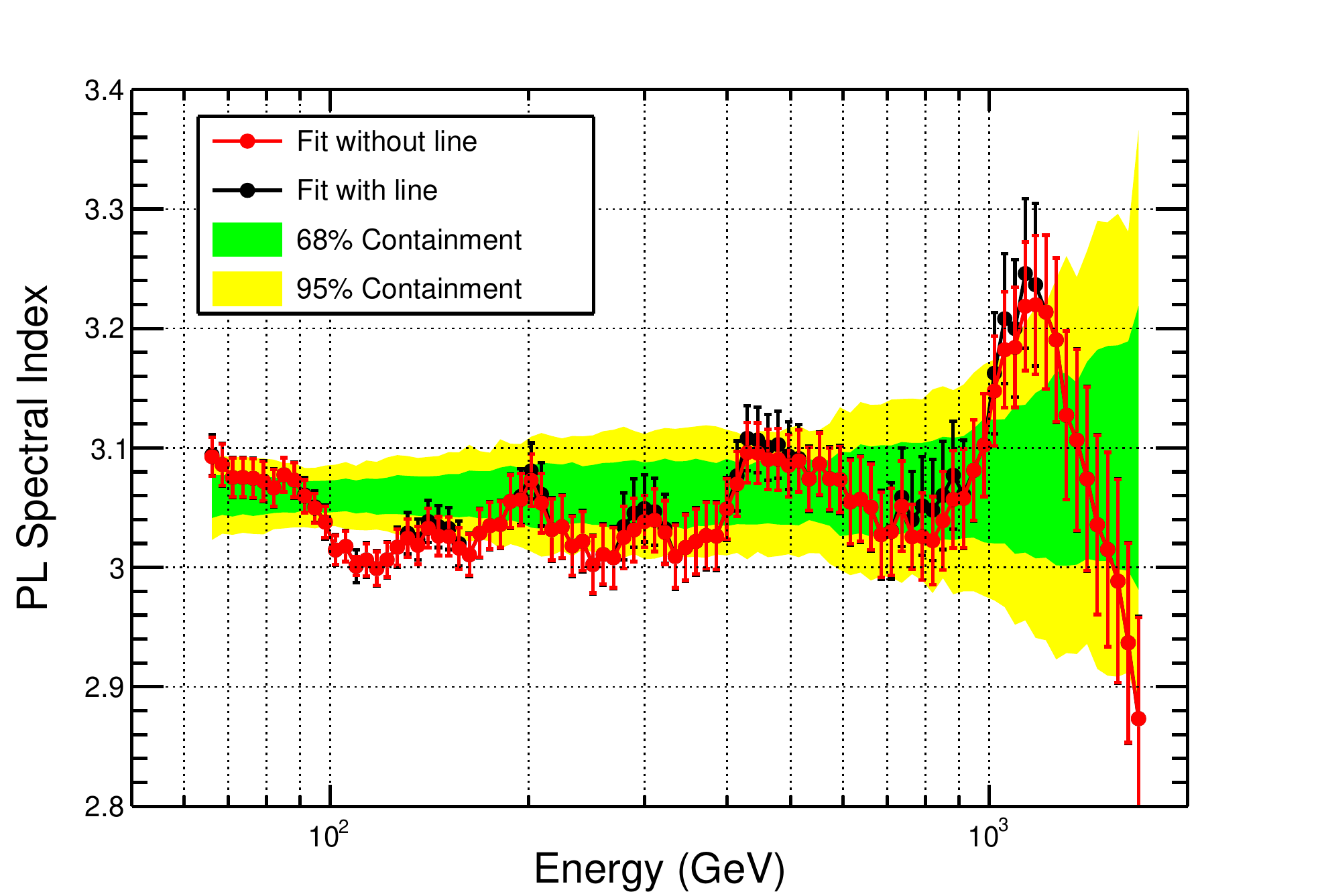} &
\includegraphics[width=\columnwidth,height=0.22\textheight,clip]{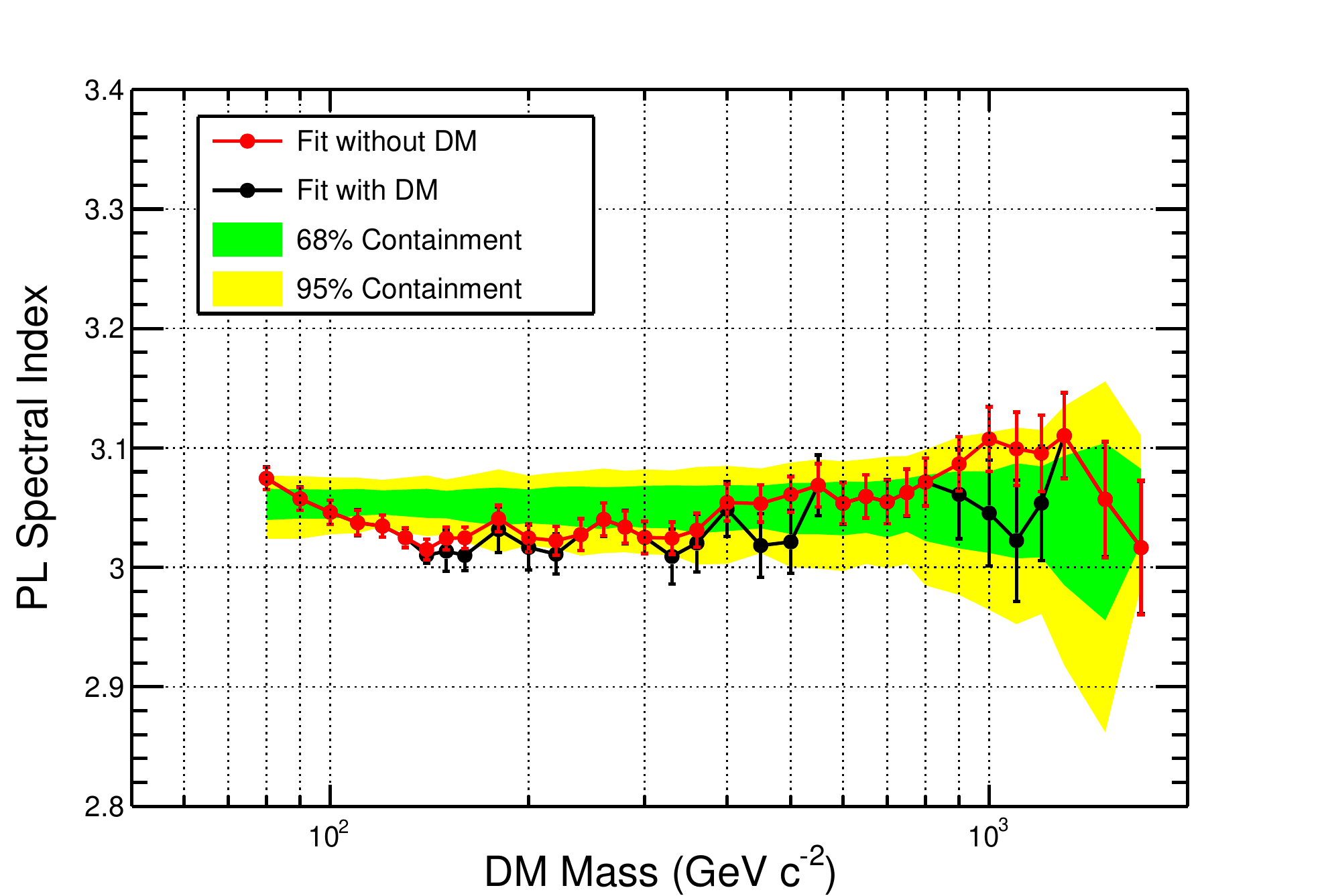}  \\
\includegraphics[width=\columnwidth,height=0.22\textheight,clip]{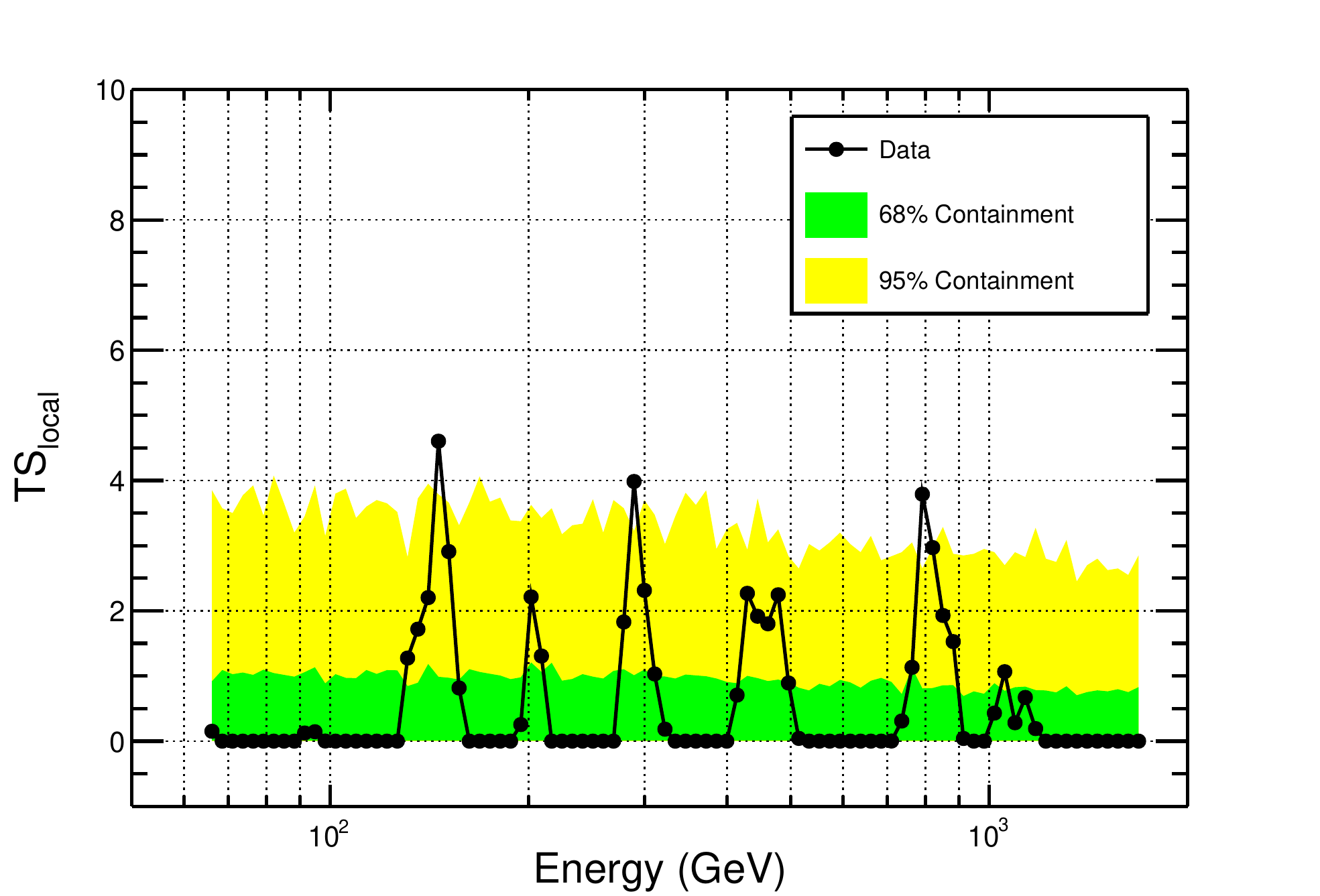} &
\includegraphics[width=\columnwidth,height=0.22\textheight,clip]{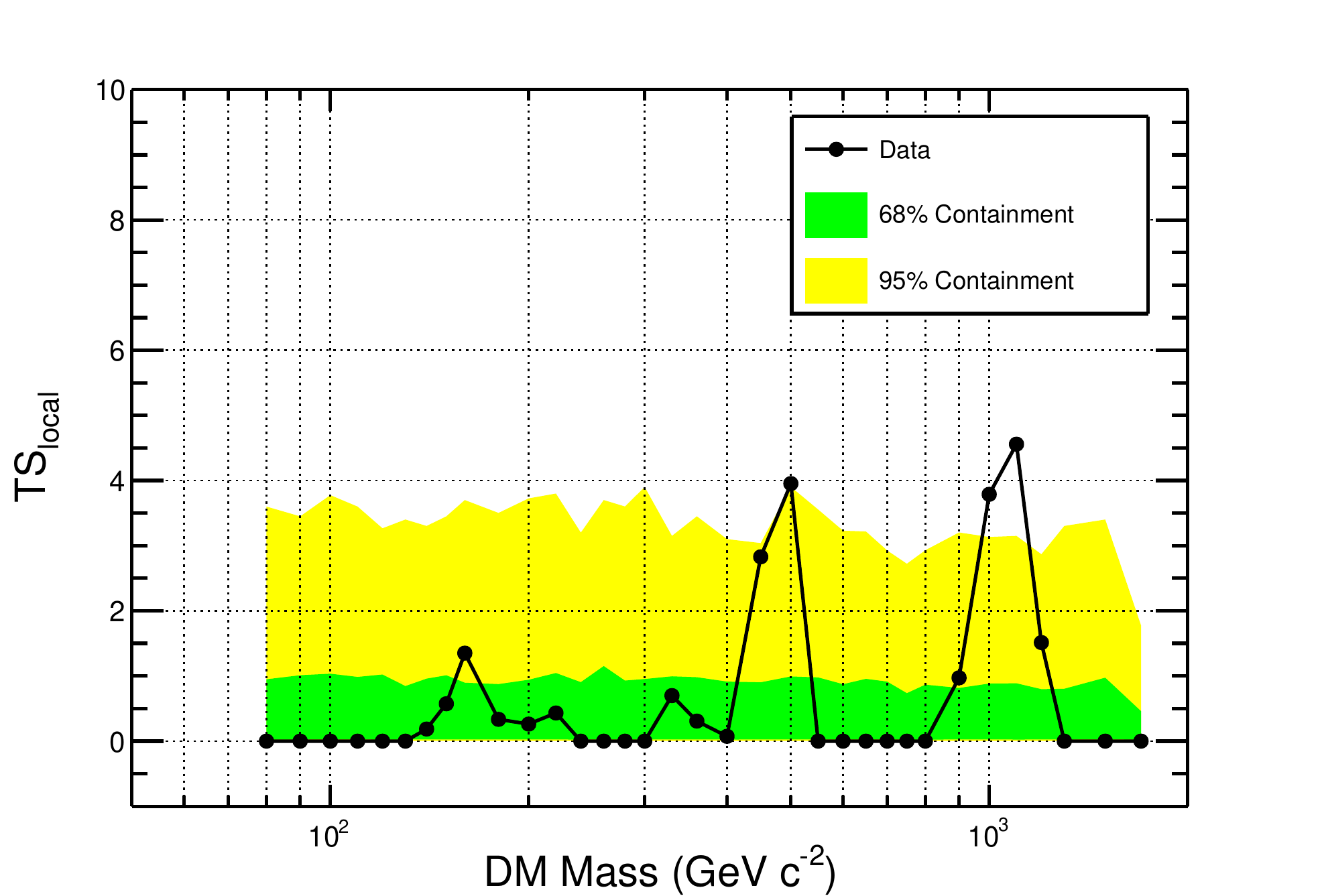} 
\end{tabular}
\caption{ 
Summary of the fit results.
The left plots have been obtained fitting the CRE 
spectrum with a delta-like line feature on top of a
PL spectrum; the right plots have been obtained assuming a feature due to DM 
annihilating into CREs on top of a PL spectrum. The black points in the
top four plots indicate the values of the PL prefactor and spectral index obtained when the 
feature is included in the fit; the red points are obtained when the 
fit is performed without the feature, i.e. setting $I_{f}(E)=0$. 
The bottom plots show the values of the Test Statistics $TS_{\rm{local}}$
for the line (left) and DM (right) models (alternative hypothesis) with
respect to the simple PL model (null hypothesis). The green and yellow
bands indicate the $68\%$ and $95\%$ confidence belts evaluated with the 
pseudo-experiment technique.} 
\label{Fig:bands_1}
\end{figure*}

\section{Results and Discussion}
\label{sec:fits}

Fig.~\ref{Fig:exmfit} shows the results of the fits performed in the energy region near $1 \units{TeV}$.
The left and right plots show respectively the results obtained when the CRE spectrum is fitted  
with a delta-like line feature or with a DM feature on top of a PL spectrum. 
The figures show a comparison of the measured counts with those predicted from the fit. 
In both cases the fitted count distributions reproduce fairly well the observed ones.  
In the figures the contributions to the count spectra from the features are also shown.
The counts due to the possible feature are always less than the $10\%$ of the total
counts in each bin. 

A comparison of the fitted spectra with the data is shown 
in Fig.~\ref{Fig:results}.
The plots in the left panels show the results obtained when fitting 
the CRE count distribution with a delta-like line feature on top of a PL spectrum,
while those in the right panels show the results obtained when fitting the 
distribution with a feature due to DM annihilations in the Galaxy 
superimposed on a PL spectrum.
In both cases, the observed count spectrum is well reconstructed in all the energy windows.

A summary of the fit results is given in Fig.~\ref{Fig:bands_1}.
In the top and in the middle panels  
the values of the fitted PL prefactor ($k$) and spectral index ($\gamma$) 
are shown as a function of energy for the two spectral models 
considered in the present analysis. 
The left plots show the results obtained for the delta-like line 
feature, while those on the right show the results obtained assuming 
a feature in the CRE spectrum due to DM annihilating into CREs. 
The values of the parameters obtained in the fit are compared 
with those obtained when the fit is performed without the
feature, setting $I_{f}(E)=0$ or equivalently $s=0$ (null hypothesis).
The values of $k$ and $\gamma$ obtained in the null hypothesis are
consistent with those obtained when $s \neq 0$. This result is 
expected, since possible spectral features are expected to be tiny. 
The plots in Fig.~\ref{Fig:bands_1} also show the confidence belts
evaluated with the pseudo-experiment technique described in Sec.~\ref{anamet}.
In most cases the fitted parameters lie within the central $95\%$ confidence 
belt.

As mentioned in Sec.~\ref{anamet}, to evaluate the local significance 
of a possible feature one can use as a Test Statistic the
value $TS_{\rm local} = -\Delta\chi^{2} = -(\chi^{2}_{1}-\chi^{2}_{0})$
where $\chi^{2}_{1}$ and $\chi^{2}_{0}$ are respectively 
the $\chi^2$ values obtained when fitting the data with 
the alternative hypothesis (line or DM signal superimposed to
the PL spectrum) and with the null hypothesis (PL spectrum).
The $TS_{\rm local}$ defined in this way is expected to obey a $\chi^{2}$
distribution with one degree of freedom since the two models differ by one free parameter.
The local significance in $\sigma$ units can be then 
evaluated as $s_{\rm{local}}=\sqrt{TS_{\rm{local}}}$.

\begin{figure*}[!htb]
\begin{tabular}{cc}
\includegraphics[width=\columnwidth,height=0.22\textheight,clip]{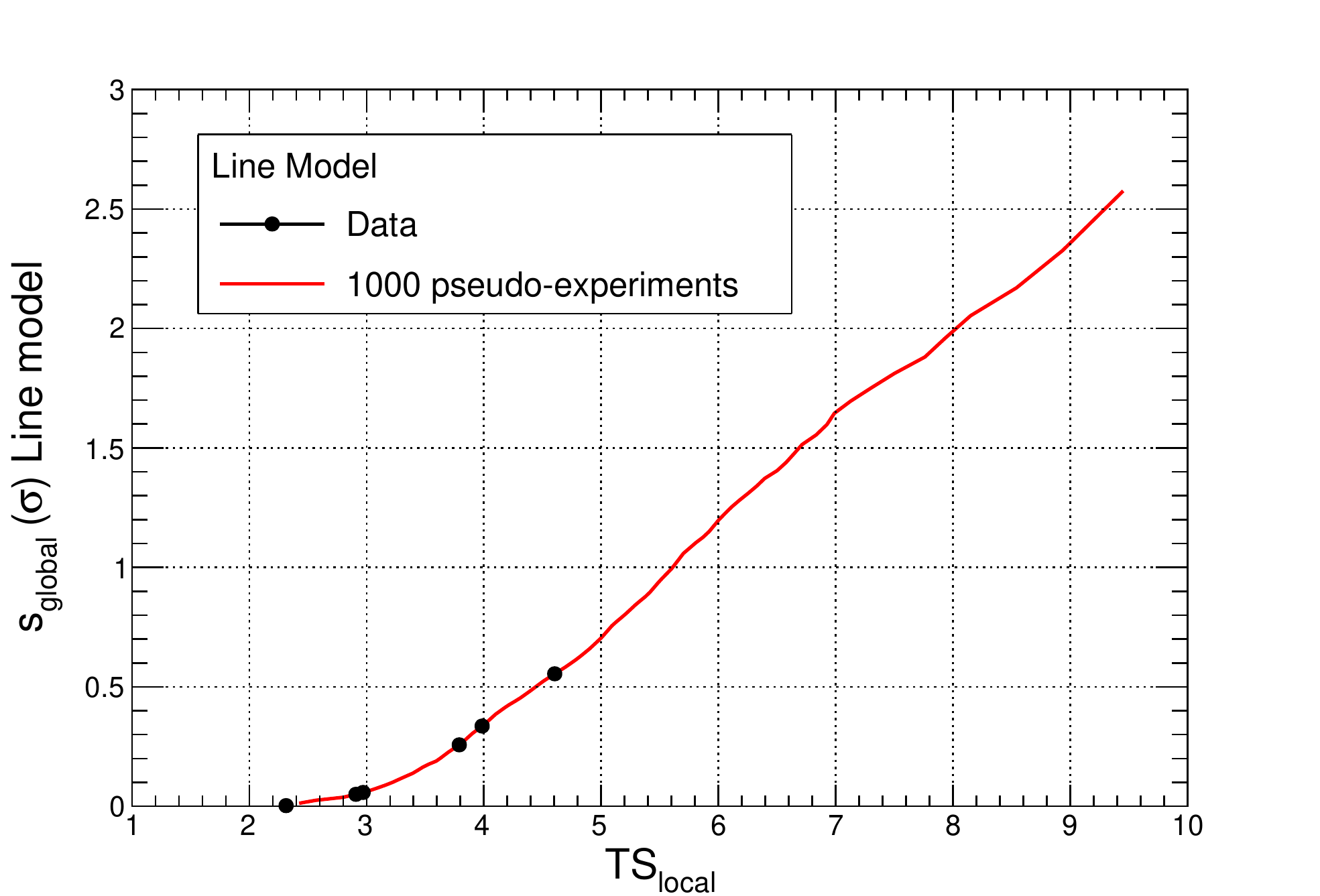} &
\includegraphics[width=\columnwidth,height=0.22\textheight,clip]{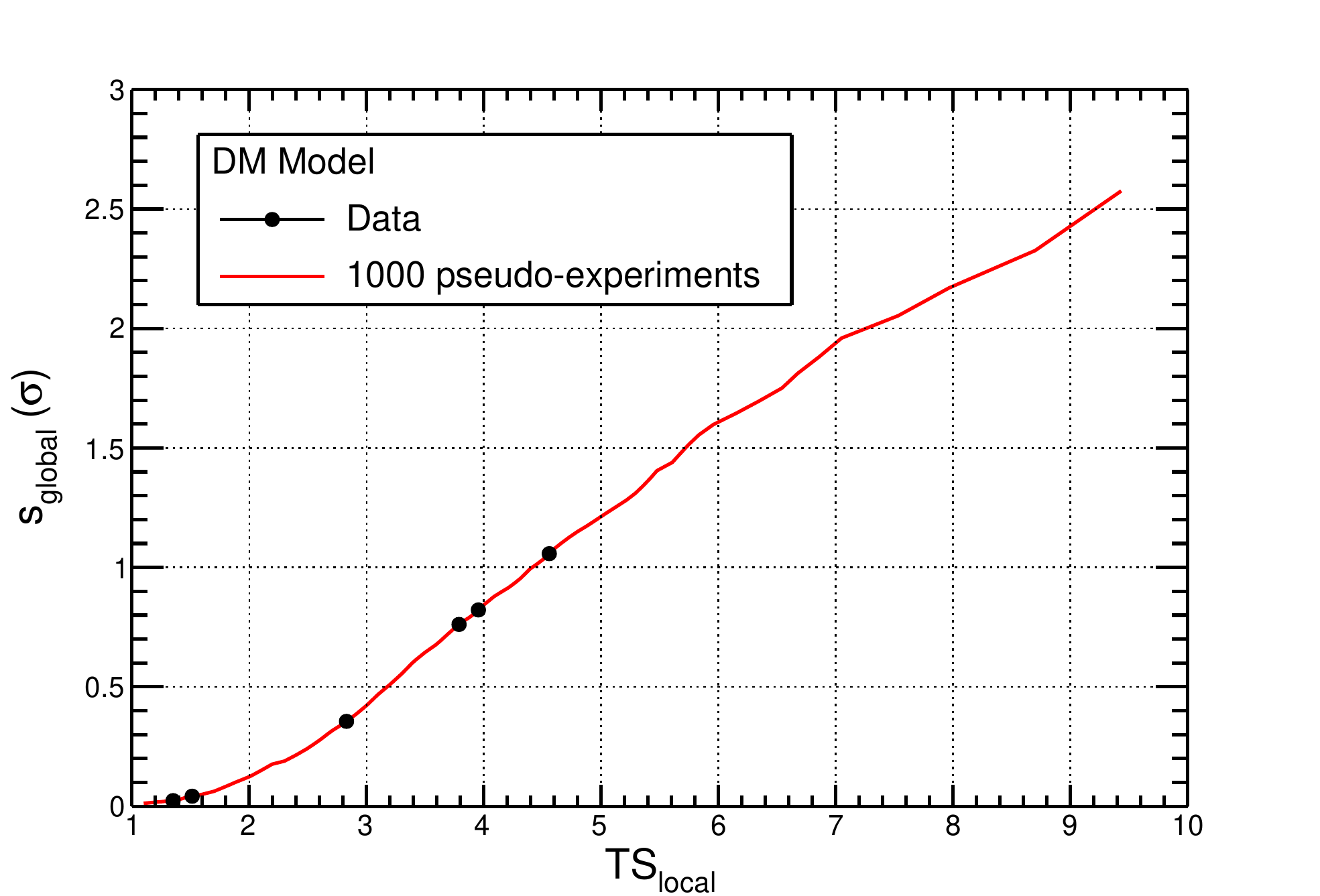} 
\end{tabular}
\caption{Conversion from $TS_{\rm{local}}$ to $s_{\rm global}$ derived from the analysis of 
the 1000 pseudo-experiments. Left panel: line model.
Right panel: DM model. The markers show some of most statistically significant local features.} 
\label{Fig:Sglob}
\end{figure*}

The bottom panels of Fig.~\ref{Fig:bands_1} show the values
of $TS_{\rm local}$ as a function of energy, compared
with the $68\%$ and $95\%$ expectation bands obtained with the
pseudo-experiment technique. In most energy windows, the values
of $TS_{\rm local}$ are close to zero and lie within the $95\%$ 
expectation band. There are some fits yielding values 
of $TS_{\rm local}$ slightly above the $95\%$ expectation bands. 
However, in the evaluation of the global significance of these
possible features, it should be kept in mind that the fits are
not independent and the number of trials should be taken into account.
As a consequence, possible features associated with a local significance larger than
$2 \sigma$ turn out to be globally insignificant.
The local significance of a 
possible feature has been evaluated from $TS_{\rm local}$.
However, since we perform many fits, to obtain the global significance 
$s_{\rm global}$, the local significance must be corrected
taking into account the effective number of trials.
For the line search we performed $88$ fits, while for the DM search we performed $32$ 
fits, but all these fits are not independent since they largely overlap in energy.

To calculate the global significances we use the 1000 pseudo-experiments 
discussed in Sec.~\ref{anamet}. For each pseudo-experiment (which corresponds to
a simulation of one full search across the entire energy range) we record 
the largest value of the local Test Statistic, $TS_{\rm max}$. 
We then calculate the quantiles of the distribution of $TS_{\rm max}$
and we evaluate the corresponding values of the global significance 
$s_{\rm global}$ assuming that $s_{\rm global}$ obeys a a half-normal distribution.

Fig.~\ref{Fig:Sglob} shows the conversion from $TS_{\rm local}$ to $s_{\rm global}$
for the line (left panel) and DM (right panel) models. 
The most significant features have global significances of 0.56$\sigma$ ($E=145\units{GeV}$) 
and 1.14$\sigma$  ($m_{DM}=1.1\units{TeV/c^{2}}$) for the line and DM model respectively.

\begin{figure*}[!ht]
\includegraphics[width=\columnwidth,height=0.22\textheight,clip]{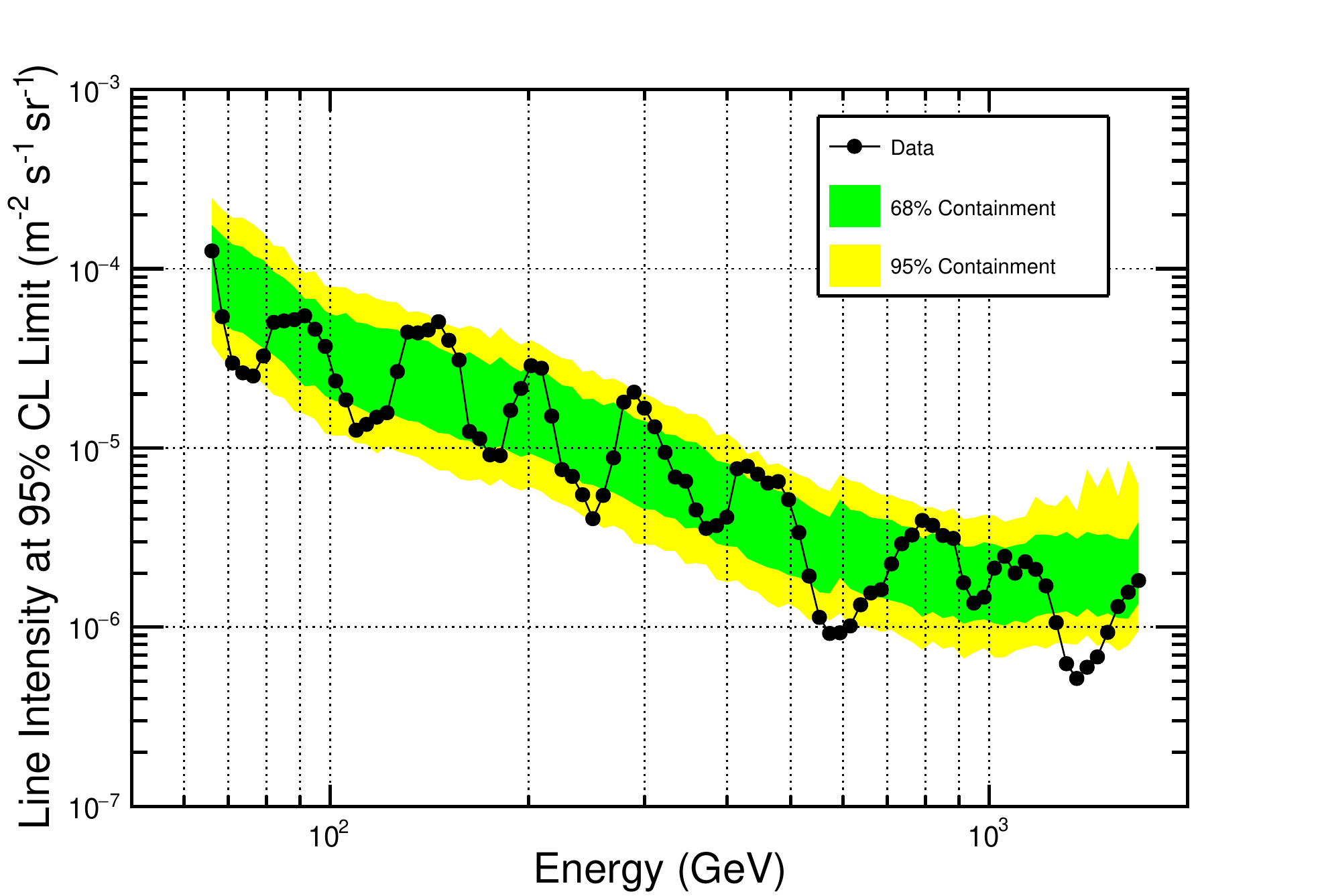} 
\includegraphics[width=\columnwidth,height=0.22\textheight,clip]{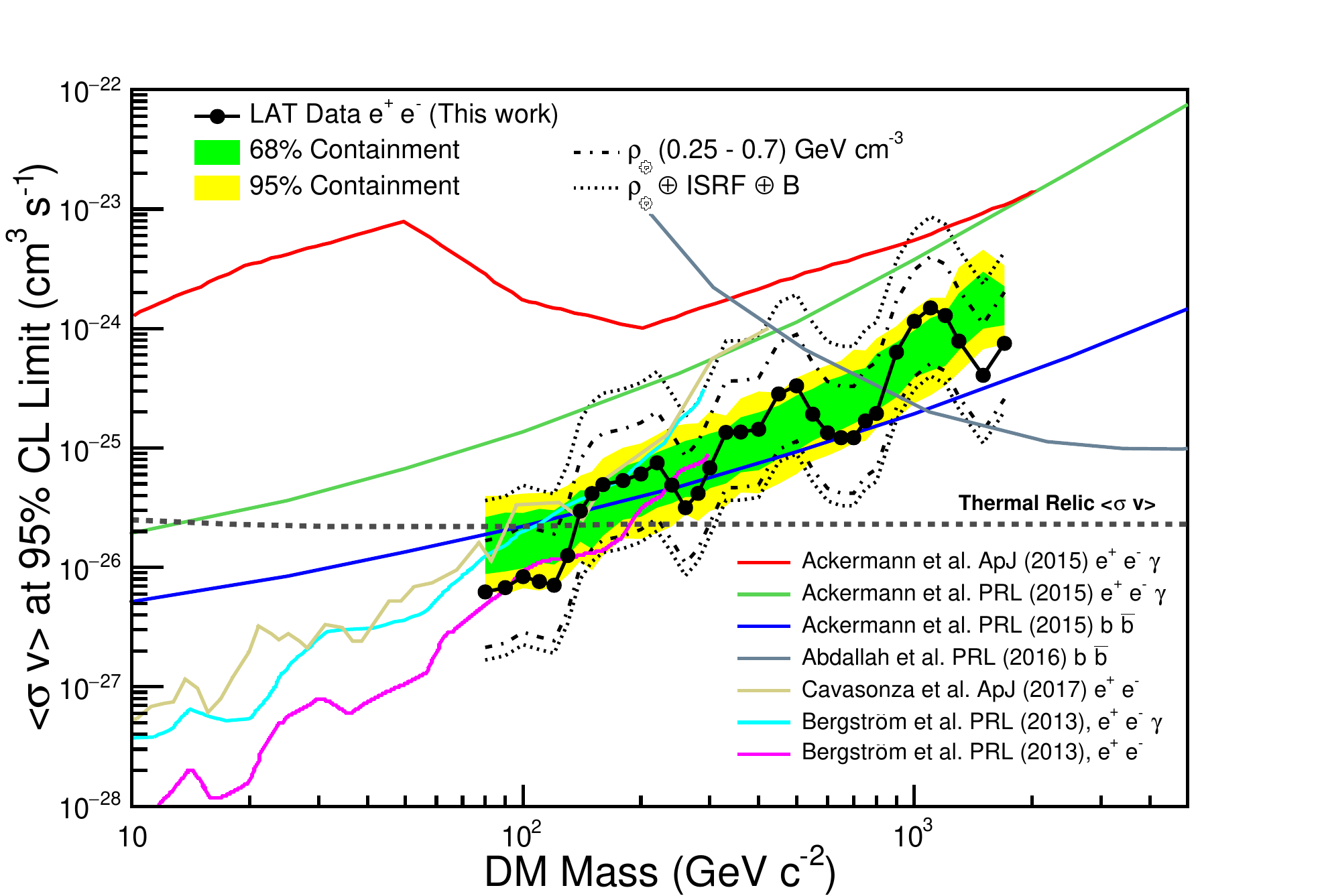} 
\caption{Upper limits on the CRE spectral features. The left plot has been obtained assuming 
a delta-like line feature on top of a power-law spectrum; the right plot 
has been obtained assuming a feature due to DM annihilating into CREs on top of a power-law spectrum. 
The plots show the upper limits at $95\%$ CL on the parameter 
describing the feature ($s$ for the line, $\langle \rm{\sigma v} \rangle$ for the DM signal). 
The dashed-dotted and the dotted black lines in the right plot indicate the variations of the limits 
on $\langle \rm{\sigma v} \rangle$ for reasonable variations of the local DM density,
of the ISRF and of the GMF (see discussion in the text).
The green and yellow $68\%$ and $95\%$ confidence belts have been evaluated with the pseudo-experiment 
technique. The dashed line in the right plot
indicates the thermal relic cross section from Steigman et al.~\cite{Steigman:2012nb}.
The colored lines indicate the upper limits on $\langle \rm{\sigma v} \rangle$
taken from Refs.~\cite{Ackermann:2015zua,Ackermann:2015fdi,Bergstrom:2013jra,Cavasonza:2016qem,Abdallah:2016ygi}.}
\label{Fig:bands}
\end{figure*}

Fig.~\ref{Fig:bands} shows the upper limits at $95\%$ confidence 
level on the parameters describing the feature
($s$ for the line, $\langle \rm{\sigma v} \rangle$ for the DM signal).
In the right panel of Fig.~\ref{Fig:bands} we also include 
dashed-dotted and dotted black lines
showing variations of the limits on $\langle \rm{\sigma v} \rangle$ 
assuming that $\rho_\odot$ can vary in the range $(0.25-0.7)\units{GeV~cm^{-3}}$
and that the ISRF together with the GMF can vary by $\pm 50\%$, respectively.   
The green and yellow bands show the $68\%$ and $95\%$ CL expectation bands, respectively, 
calculated from the pseudo-experiments
discussed in Sec.~\ref{anamet}. 
Since the limits lie within the $95\%$ CL expectation bands, the Fermi LAT data do not provide
evidence of any feature at the $2 \sigma$ (local) level, either in the case of a 
delta-like line or in the case of a signal from DM annihilations in the Galaxy.

As shown in Fig.~\ref{FigDMCRE}, the differences among the CRE spectra 
measured by DAMPE, CALET, AMS-02 and the Fermi LAT are within $20\%$ in 
the \units{TeV} region. 
Assuming this uncertainty on the CRE spectrum, this would imply a variation of the current upper limits 
at most at the same level, which is significantly smaller than the variations originating from the uncertainties 
in the DM models, due to for instance the uncertainties on the local DM density, or to those on the GMF and on the ISRF.
Although our analysis of the LAT data accounts for systematic uncertainties, in the most conservative interpretation,
it could be argued that the 20$\%$ differences represent the limit to what could be resolved. 

The $95\%$ upper limits on the
velocity-averaged DM annihilation cross section $\langle \rm{\sigma v} \rangle$ into $e^{+}e^{-}$ 
pairs obtained in the present analysis lie below the thermal relic cross section
calculated by Steigman et al.~\cite{Steigman:2012nb} for DM masses up to 
$100 \units{GeV/c^2}$. 
Our limits are consistent with those obtained by Bergstr\"om et al. in Ref.~\cite{Bergstrom:2013jra}
when studying the AMS-02 data on the positron fraction~\cite{Aguilar:2013qda}
and by Cavasonza et al. in Ref.~\cite{Cavasonza:2016qem} when studying 
the AMS-02 electron and positron data~\cite{Aguilar:2014mma}
in the range around $100 \units{GeV}$ where they overlap in energy.\footnote{In 
Refs.~\cite{Bergstrom:2013jra} and ~\cite{Cavasonza:2016qem} the local DM density was assumed to be
$0.4 \units{GeV~cm^{-3}}$ and $0.3 \units{GeV~cm^{-3}}$ respectively.}

The present limits are also competitive with those 
obtained by the Fermi LAT Collaboration when studying the gamma-ray emission 
from the Virgo Galaxy Cluster~\cite{Ackermann:2015fdi} 
and from the Milky Way dwarf spheroidal Galaxies~\cite{Ackermann:2013yva,Ackermann:2015zua}
in the channel $e^{+}e^{-} \gamma$
and are similar to those in the $b \bar{b}$ channel~\cite{Ackermann:2013yva,Ackermann:2015zua}.
Finally, our limits are consistent with the limits obtained from the 
analysis of the gamma rays from the inner Galactic halo performed by the H.E.S.S.~Collaboration 
assuming a cuspy DM profile~\cite{Abdallah:2016ygi}.

\section{Summary}
\label{sec:summ}

In this work, we have used the Fermi-LAT CRE data to
search for possible features in the spectrum originating from the 
direct annihilation of DM particles into $e^{+}e^{-}$ pairs in the Galaxy halo.
We searched for spectral features from $42 \units{GeV}$ to $2 \units{TeV}$,
thus extending the previous results based on the AMS-02 electron-positron data in the
energy range above $300 \units{GeV}$~\cite{Bergstrom:2013jra,Cavasonza:2016qem}.
The current results have been also compared with the constraints based on the DM annihilation to gamma rays.

The current analysis yields no evidence for a line or a DM feature.
With the DM model assumed in the present analysis or for a pure line case, 
we do not find any indication for the presence of a feature at $1.4\units{TeV}$, as suggested 
by the recent DAMPE measurements~\cite{Chen:2017tva,Cao:2017sju,Fowlie:2017fya,Athron:2017drj}. 

The limits on the intensity of a line-like feature can be used, in principle,
to study other DM models which also produce a feature in the spectrum.
In this case, from an approximate  match of the DM feature with the line,
constraints on the DM model can be derived.
Similarly, they can also be used to derive 
constraints on the presence of nearby CRE accelerators, like pulsars
or supernova remnants.
A quantitative analysis lies, however, beyond the scope
of the present work.

Limits in the case of decaying DM with mass $2m$ can be easily obtained from the case of
annihilating DM of mass $m$ with the simple transformation: 
$\Gamma = 1/2 \langle \rm{\sigma v} \rangle \rho_\odot/m$,  where $\Gamma$
is the DM decay rate. We have explicitly checked that this approximation is valid
at a few percent level or less.

\begin{acknowledgments} 
The Fermi LAT Collaboration acknowledges generous ongoing support
from a number of agencies and institutes that have supported both the
development and the operation of the LAT as well as scientific data analysis.
These include the National Aeronautics and Space Administration and the
Department of Energy in the United States, the Commissariat \`a l'Energie Atomique
and the Centre National de la Recherche Scientifique / Institut National de Physique
Nucl\'eaire et de Physique des Particules in France, the Agenzia Spaziale Italiana
and the Istituto Nazionale di Fisica Nucleare in Italy, the Ministry of Education,
Culture, Sports, Science and Technology (MEXT), High Energy Accelerator Research
Organization (KEK) and Japan Aerospace Exploration Agency (JAXA) in Japan, and
the K.~A.~Wallenberg Foundation, the Swedish Research Council and the
Swedish National Space Board in Sweden.
 
Additional support for science analysis during the operations phase is gratefully
acknowledged from the Istituto Nazionale di Astrofisica in Italy and the Centre
National d'\'Etudes Spatiales in France. This work performed in part under DOE
Contract DE-AC02-76SF00515.
\end{acknowledgments}

\appendix

\begin{figure*}[!htb]
\begin{tabular}{cc}
\includegraphics[width=\columnwidth,height=0.22\textheight,clip]{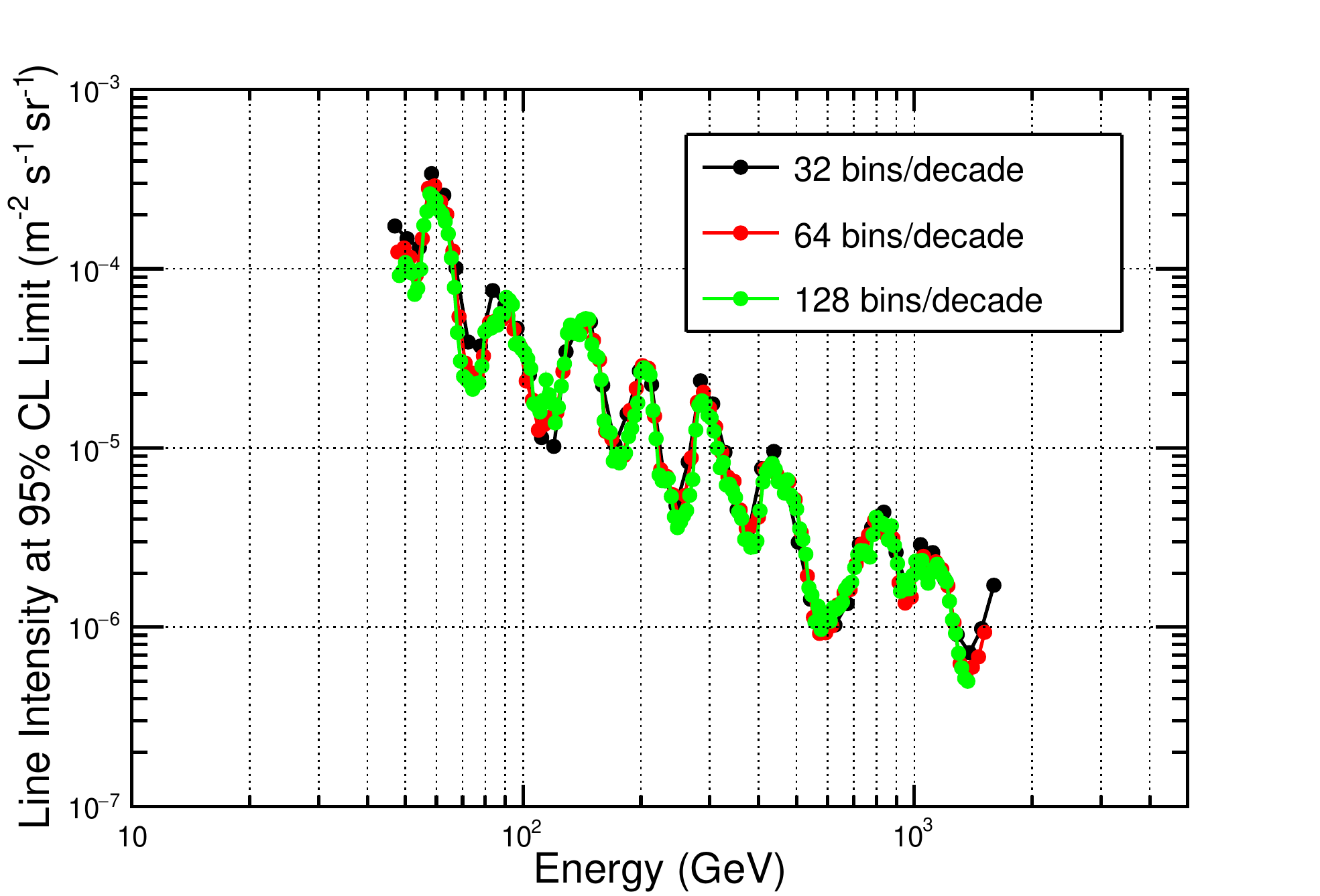} &
\includegraphics[width=\columnwidth,height=0.22\textheight,clip]{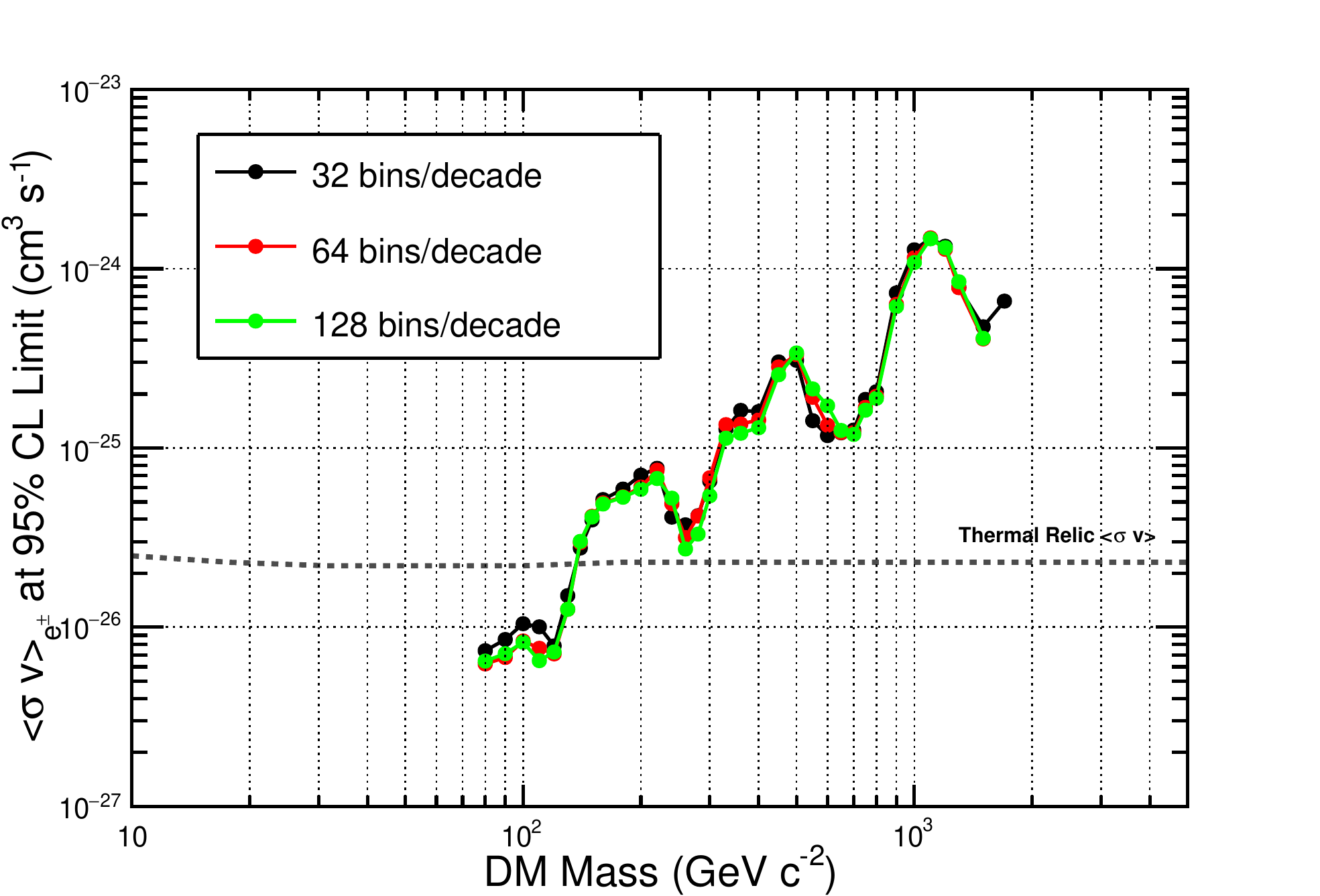} \\
\end{tabular}
\caption{Comparison of the upper limits at $95\%$ confidence level on the line intensity 
(left plot) and on the velocity averaged cross section $\langle \rm{\sigma v} \rangle$ (right plots) obtained with 
$32$ (black symbols), $64$ (red symbols) and $128$ (green symbols) energy bins per decade.
The half-width of the fit windows is $0.35 E_{line}$ for the line fits and $0.5 m_{DM}$ for the DM fits.} 
\label{Fig:UL_bins}
\end{figure*}

\begin{figure*}[!htb]
\begin{tabular}{cc}
\includegraphics[width=\columnwidth,height=0.22\textheight,clip]{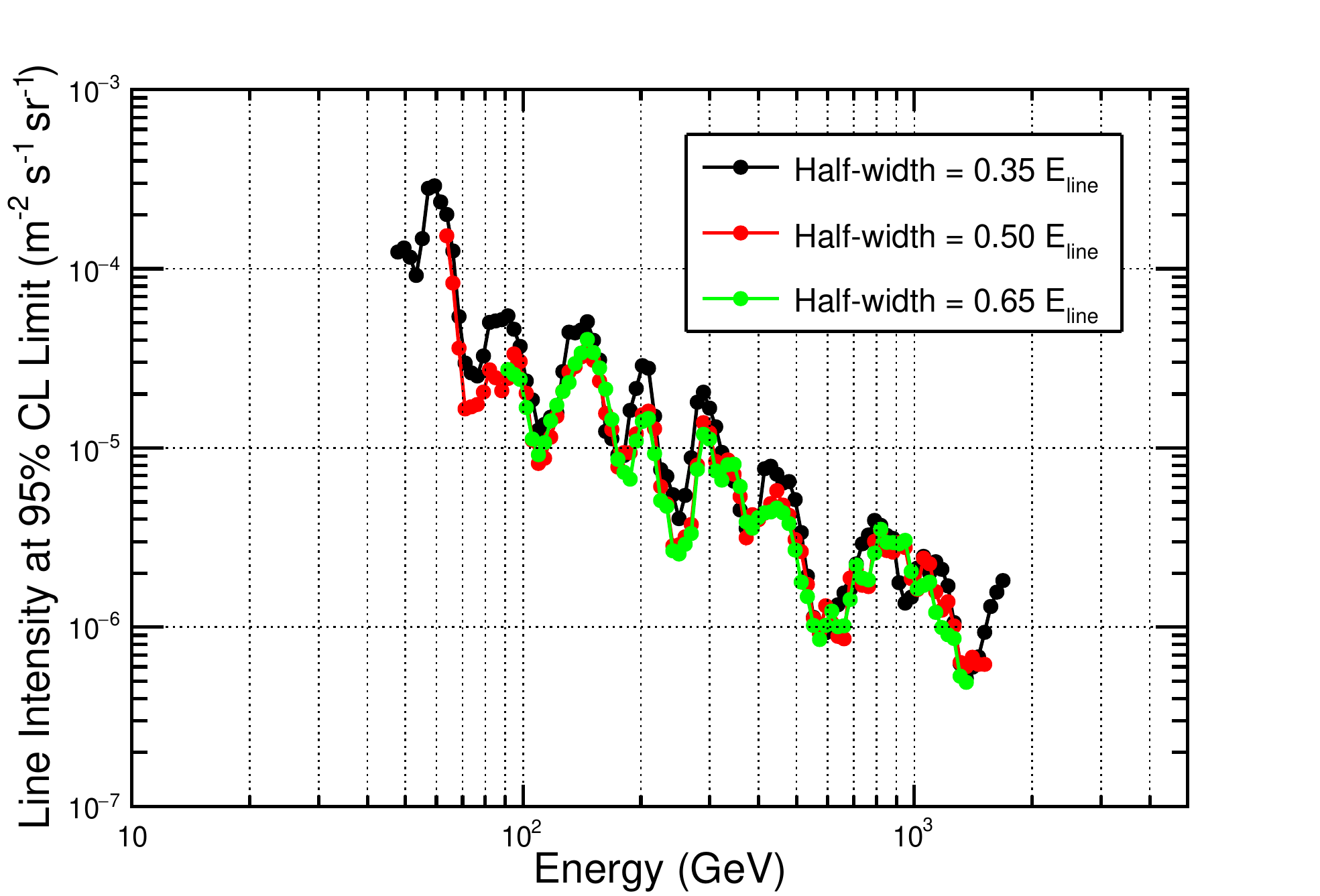} &
\includegraphics[width=\columnwidth,height=0.22\textheight,clip]{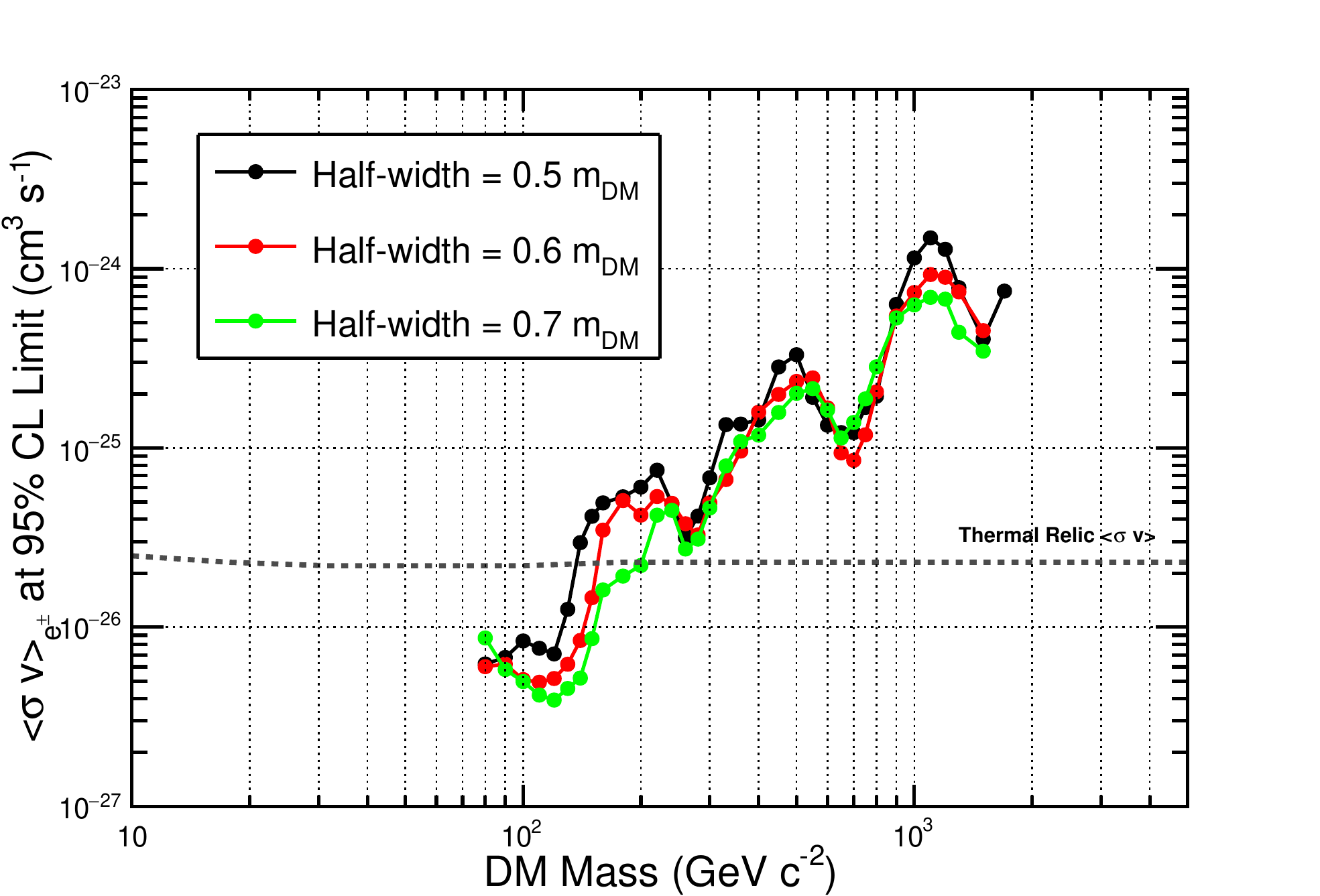} \\
\end{tabular}
\caption{Comparison of the upper limits at $95\%$ confidence level on the line intensity 
(left plot) and on the velocity averaged cross section $\langle \rm{\sigma v} \rangle$ (right plots) obtained with 
fit windows of different widths. The fits have been performed dividing the whole energy interval in $64$ bins per decade. } 
\label{Fig:UL_width}
\end{figure*}

\section {Effects of the energy binning and of the window size on the constraints}
\label{sec:binning}
To test the reliability of our analysis method, we have also studied the 
dependence of the fit results on the energy binning and on the choice of the window size. 
Fig.~\ref{Fig:UL_bins} shows a comparison of the upper limits on the line
intensity and on the DM velocity averaged cross section $\langle \rm{\sigma v} \rangle$ obtained
dividing the energy interval in $32$, $64$ and $128$ bins per decade
and assuming the nominal sizes for the fit windows (i.e. $0.35 E_{line}$ and $0.5 m_{DM}$).
As can be seen from the figure, in both cases the upper limits are
almost independent of the energy binning. 

In Fig.~\ref{Fig:UL_width} we compare the upper limits on the line and  
on the DM annihilation cross sections obtained dividing the energy
interval in $64$ bins per decade and assuming different sizes for the 
fit windows.
In the case of the line fit, the upper limits are almost independent on 
the window size. The choice of the window size determines 
the interval $[E_{l1},E_{l2}]$ of possible line energies, since the conditions
$E_{l1}-\Delta E_{l1} \geq 42 \units{GeV}$ and $E_{l2}+\Delta E_{l2} \leq 2 \units{TeV}$
have to be satisfied, and larger window sizes will result in smaller energy intervals. 
Since the obtained line intensity is found to be independent of the chosen window size, 
we choose the smallest possible window size, $0.35 E_{line}$. 
Smaller windows are not appropriate for this analysis since, due to the
energy resolution ($15\% - 35\%$), the line features are expected 
to spread over several energy bins, which should all be included in the
fit windows. Likewise, for the DM fit, 
the obtained upper limits on the DM-induced flux show a mild dependence on the  
window size. In our analysis we choose a window size of $0.5 m_{DM}$ 
since it provides the most conservative limits. 
Also in this case smaller windows are not appropriate because 
the feature is expected to spread over many energy bins.


\bibliographystyle{apsrev4-1}
\bibliography{LATCREDM.bib}{}

\end{document}